\documentclass{IEEEoj}

\ifCLASSINFOpdf
 \usepackage[pdftex]{graphicx}
\else
 \usepackage[dvips]{graphicx}
\fi
\usepackage{cite}
\usepackage{amsmath,amssymb,amsfonts,lipsum}
\usepackage{float}

\usepackage{algorithmic}
\usepackage{algorithm}
\usepackage{stfloats}
\usepackage{graphicx}
\usepackage{ragged2e}
\usepackage{textcomp}
\usepackage{multirow}
\usepackage{color}
\usepackage{xcolor}
\usepackage{booktabs}
\def\BibTeX{{\rm B\kern-.05em{\sc i\kern-.025em b}\kern-.08em
    T\kern-.1667em\lower.7ex\hbox{E}\kern-.125emX}}

\usepackage[caption=false,font=normalsize,labelfont=sf,textfont=sf]{subfig}
\def\BibTeX{{\rm B\kern-.05em{\sc i\kern-.025em b}\kern-.08em
    T\kern-.1667em\lower.7ex\hbox{E}\kern-.125emX}}
\AtBeginDocument{\definecolor{ojcolor}{cmyk}{0.93,0.59,0.15,0.02}}
\def\OJlogo{\vspace{-14pt}\includegraphics[height=28pt]{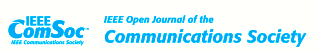}}

\begin{document}
\receiveddate{XX Month, XXXX}
\reviseddate{XX Month, XXXX}
\accepteddate{XX Month, XXXX}
\publisheddate{XX Month, XXXX}
\currentdate{XX Month, XXXX}
\doiinfo{OJCOMS.XXXX.XXXXXXX}
\title{Energy-Efficient Clustered Cell-Free Networking with Access Point Selection}
\author{OUYANG ZHOU\authorrefmark{1}, STUDENT MEMBER, IEEE, JUNYUAN WANG\authorrefmark{1}\authorrefmark{2}, MEMBER, IEEE, FUQIANG LIU\authorrefmark{1}\authorrefmark{3}, MEMBER, IEEE AND JIANGZHOU WANG\authorrefmark{4}, FELLOW, IEEE}
\affil{College of Electronic and Information Engineering, Tongji University, Shanghai, 201804, China}
\affil{Institute of Advanced Study, Tongji University, Shanghai, 200092, China}
\affil{College of Computer Science and Technology, Huaibei Normal University, Huaibei, 235000, China}
\affil{School of Engineering, University of Kent, Canterbury, CT2 7NZ, U.K.}
\corresp{CORRESPONDING AUTHOR: Junyuan Wang (e-mail: junyuanwang@tongji.edu.cn).}
\authornote{This paper was presented in part in IEEE International Symposium on Information Theory (ISIT), Espoo, Finland, June 2022 \cite{Conference}. \\
This work was supported in part by the National Nature
Science Foundation of China under Grant 62371344 and 62001330,
and in part by the Fundamental Research Funds for the Central
Universities.}
\markboth{Energy-Efficient Clustered Cell-Free Networking with Access Point Selection}{ZHOU \textit{et al.}}

\begin{abstract}
Ultra-densely deploying access points (APs) to support the increasing data traffic would significantly escalate the cell-edge problem resulting from traditional cellular networks. By removing the cell boundaries and coordinating all APs for joint transmission, the cell-edge problem can be alleviated, which in turn leads to unaffordable system complexity and channel measurement overhead. A new scalable clustered cell-free network architecture has been proposed recently, under which the large-scale network is flexibly partitioned into a set of independent subnetworks operating parallelly. In this paper, we study the energy-efficient clustered cell-free networking problem with AP selection. Specifically, we propose a user-centric ratio-fixed AP-selection based clustering (UCR-ApSel) algorithm to form subnetworks dynamically. Following this, we analyze the average energy efficiency achieved with the proposed UCR-ApSel scheme theoretically and derive an effective closed-form upper-bound. Based on the analytical upper-bound expression, the optimal AP-selection ratio that maximizes the average energy efficiency is further derived as a simple explicit function of the total number of APs and the number of subnetworks. Simulation results demonstrate the effectiveness of the
derived optimal AP-selection ratio and show that the proposed UCR-ApSel algorithm
with the optimal AP-selection ratio achieves around $40\%$ higher energy efficiency than
the baselines. The analysis provides important insights to the design and optimization of
future ultra-dense wireless communication systems.
\end{abstract}

\maketitle

\section{Introduction}
Traditional cellular network partitions the wireless network into a set of cells, and one base-station (BS) is located at the cell center to serve users within its coverage. Since users are randomly distributed, there might be users located in the cell boundary areas, leading to the well-known cell-edge problem, that is, the cell-edge users experience poor network services due to low received signal power and strong inter-cell interference. To support the extremely high data rate and low traffic delay of numerous intelligent devices and services, such as autonomous vehicles, smart homes, digital twins and holographic communications, BSs would be ultra-densely deployed in the sixth-generation (6G) mobile communication systems \cite{9349624}, which in turn exacerbates the cell-edge problem. Therefore, the cellular network architecture might not be suitable for future ultra-dense wireless communication systems.

To avoid the cell-edge problem, coordinating all the geographically distributed access points (APs), which are equivalent to BSs, to jointly serve all users on the same time-frequency resources has been researched in distributed antenna systems (DASs) \cite{1545835, 4114255, 7031453}, network multiple input multiple output (MIMO) \cite{6095627}, cloud radio access network (C-RAN) \cite{pan} and cell-free massive MIMO \cite{7827017, 9650567}.  It has been demonstrated that such a fully cooperative network achieves high data rate and high energy efficiency \cite{5770665, 7031453, 6502484}. However, coordinating hundreds or even thousands of APs to serve users leads to prohibitively huge channel measurement overhead, significantly high signal processing complexity and extremely stringent backhaul requirement, which are unaffordable in practice. To reduce the joint transmission complexity and improve the system scalability, it was proposed to serve each user only by its nearby APs that determine its channel \cite{1545835, 6783666, 6827165, 8244310}. Yet the serving AP sets of different users might overlap, i.e., sharing some APs, making the signal processing of different users' serving AP sets coupled with each other and hence difficult to be optimized. Moreover, although serving each user by its nearby AP set could avoid the cell-edge problem, the independent transmissions among different users' AP sets could lead to severe inter-user interference, especially when two users' AP sets are largely overlapping. This greatly degrades system spectral efficiency and user experience. 

\subsection{Clustered Cell-Free Networking}
In order to facilitate joint processing of APs to alleviate the cell-edge problem with tolerable complexity as well as avoiding strong inter-user interference, \cite{JunyuanTWC, 8007415, 2022} proposed a novel clustered cell-free network architecture, with which a large-scale wireless network is flexibly partitioned into a set of small nonoverlapping subnetworks such that the users strongly interfering with each other are assigned into the same subnetwork, and little interference exists among different subnetworks. The subnetworks can then operate in parallel, and joint processing can be just performed inside each subnetwork independently to cancel out the intra-subnetwork interference. As each parallel operating subnetwork contains a small amount of APs and users, the joint processing complexity and the channel state information (CSI) feedback can be kept at a relatively low level.

A key problem arising from such a clustered cell-free network architecture is how to group APs and users to form subnetworks, which is referred to as clustered cell-free networking. \cite{4907459, 6376184, 7738565, 8849663} proposed AP-centric clustering strategies, which first cluster APs and then assign each user to the cluster containing its associated AP. The subnetworks can be also formed in a user-centric way \cite{6756959, 7317799, 8644255, 9343832}. In \cite{6756959, 7317799}, user-centric virtual cell is first constructed for each user  by selecting its closest APs, and then the virtual cells are merged as a subnetwork if their mutual interference is strong \cite{6756959}  or they are overlapping with each other \cite{7317799}, which, however, usually leads to unbalanced subnetworks with a giant cluster as depicted in \cite{JunyuanTWC, 8007415, 2022}. \cite{8644255, 9343832} proposed to group users into clusters first and then assign APs to the clusters according to the large-scale fading coefficients. Whether AP-centric or user-centric clustering should be adopted for clustered cell-free networking and which clustering algorithm leads to the best performance need to be investigated.

\subsection{AP Selection for Improving Energy Efficiency}
During the commercialization of the fifth-generation (5G) mobile communication systems, the extremely huge energy consumption resulting from densely deployed APs turns out to be a major concern to network operators \cite{9444152}. It is essential to design an energy-efficient networking scheme for future ultra-dense wireless communication systems. In fact, there is no need to involve the APs far from users in signal transmission, which only contribute marginal gains to the desired signal power yet cause loads of energy consumption and inter-user interference. Therefore, AP selection has been believed to be a feasible solution to energy saving when APs are ultra-densely deployed in the future. A number of research works proposed to select a small number of APs for each user according to the large-scale fading coefficients \cite{8000355}, the received signal power \cite{9444152} or the channel capacity \cite{8234671}. However, selecting APs for each user independently could activate an AP for serving only one single user, resulting in low energy efficiency. Joint AP selection and resource allocation was thus studied to reduce the power consumption \cite{9148948} or maximize the energy efficiency \cite{8695055, 9110914}. However, the existing optimal or suboptimal solutions are  usually obtained by iterative algorithms, which are of high computational complexity.

In a clustered cell-free network, as the whole network is partitioned into subnetworks dynamically according to the channel gains, the network partitioning result varies as users move. As a result, it is difficult to apply the existing joint AP selection and power allocation algorithms in practical clustered cell-free networks due to the high computational and time complexity though with superior performance. An alternative way is to incorporate a simple yet effective  AP selection algorithm into clustered cell-free networking, which is of great practical importance. As the energy consumption increases with the number of active APs, how many APs should be activated to maximize the energy efficiency is a key question, which can be addressed by analyzing the system performance theoretically.

\subsection{Related Works}
System performance analysis of cell-free massive MIMO has been thoroughly studied \cite{7031453, 9394765, 8886730} in the literature.
\cite{7031453} studied the downlink transmission and analyzed the average user rate (i.e., the achievable ergodic rate averaged over AP and user positions) with both maximum ratio transmission (MRT) and zero-forcing beamforming (ZFBF). \cite{9394765} assumed that APs are Poisson point process (PPP) distributed and formulated a lower-bound of the downlink average spectral efficiency based on stochastic geometry. With both APs and users randomly distributed, approximate expressions of the average uplink rate were derived in \cite{8886730}. However, these performance analyses only considered the case where all the APs simultaneously serve all users. With clustered cell-free networking, the inter-subnetwork interference should be taken into account when analyzing the average energy efficiency, as the whole network is partitioned into a set of parallel operating subnetworks. The authors in \cite{6756959, 9347738} studied the energy efficiency maximization problem and the sum rate maximization problem of clustered cell-free networks, respectively. Due to the lack of a simple yet effective analytical model, the inter-subnetwork interference was assumed to be neglected \cite{6756959} or
a constant \cite{9347738}. Moreover, dynamically clustering APs and users into subnetworks leads to varying network structure, making the performance analysis extremely difficult or even intractable. How to analyze the average energy efficiency theoretically for clustered cell-free networks by properly modeling the inter-subnetwork interference so as to determine the optimal number of APs that should be activated to maximize the average energy efficiency is a key challenge, which is the main focus of this paper.

\subsection{Our Contributions}
In this paper, we study the downlink transmission of a clustered cell-free network with a large amount of randomly distributed APs and users. The large-scale network is partitioned into a set of nonoverlapping subnetworks operating parallelly and ZFBF is adopted in each subnetwork for intra-subnetwork interference cancellation. The main contributions of this paper are summarized as follows:$\footnote{In
addition to
the conference version \cite{Conference} that is confined to the average per-user rate analysis, this paper further analyzes the average energy efficiency and optimizes AP selection for
average energy efficiency maximization.}$
\begin{itemize}
	\item In contrast to the existing works which assumed that all the APs are activated for signal transmission, a user-centric ratio-fixed AP-selection based clustering (UCR-ApSel) algorithm is proposed to incorporate AP activation into clustered cell-free networking while fixing the ratio of the number of APs to the number of users in each subnetwork at some constant $\lambda$ to ensure fairness. Simulation results show that the proposed UCR-ApSel algorithm improves both energy efficiency and user fairness.
    \item The average energy efficiency achieved with the proposed UCR-ApSel algorithm is analyzed and a closed-form upper-bound is derived. 
    To the best of our knowledge, this is the first work that theoretically analyzes the average energy efficiency of clustered cell-free networks over AP and user positions, which could shed light on the design of energy-efficient clustered cell-free ultra-dense networks.
     \item Based on the closed-form expression of the upper-bound, the optimal AP-selection ratio $\lambda^{*}$ maximizing the average energy efficiency is obtained as a simple explicit function of the total number of APs and the number of subnetworks. Simulations corroborate that the analytical result provides a good estimation of the optimal AP-selection ratio and the proposed UCR-ApSel algorithm with the optimal AP-selection ratio achieves significant gains over the benchmarks.
\end{itemize}

The remainder of this paper is organized as follows. Section II introduces the system model. The user-centric ratio-fixed AP-selection based clustering algorithm is proposed in Section III. Section IV presents the corresponding average energy efficiency analysis and derives the optimal AP-selection ratio. Discussions are provided in Section V, and concluding remarks are summarized in Section VI.

Throughout this paper, $*$, $T$, $\dagger$, $\mathbb{E}[\cdot]$ and $\lfloor\cdot\rfloor$ represent the conjugate, transpose, conjugate transpose, expectation and floor operators, respectively. $\|\cdot\|_{1}$ and $\|\cdot\|$ stand for the L1 norm and L2 norm of a vector. $x \sim \mathcal{C} \mathcal{N}\left(u, \sigma^{2}\right)$ represents a complex Gaussian random variable with mean $u$ and variance $\sigma^{2}$. $|\mathcal{X}|$ denotes the cardinality of set $\mathcal{X}$.
 \begin{table}[t]
     \centering
     \caption{Main Notations}
     \resizebox{\linewidth}{!}{
     \begin{tabular}{|c|c|} \hline
         $K$ & Number of users \\ \hline
         $L$ & Number of APs \\ \hline
         $\mathcal{U}=\left\{u_{1}, u_{2}, \cdots, u_{K}\right\}$ & User set \\ \hline
         $\mathcal{B}=\left\{b_{1}, b_{2}, \cdots, b_{L}\right\}$ & AP set \\ \hline
         $M$ & Number of subnetworks \\ \hline
         $\mathcal{U}_{m}$ & Set of users in the $m$th subnetwork \\ \hline
         $\mathcal{B}_{m}$ & Set of APs in the $m$th subnetwork \\ \hline
         $\mathcal{S}_{m}$ & Set of users and APs in the $m$th subnetwork \\ \hline
         $K_{m}$ & Number of users in the $m$th subnetwork \\ \hline
         $L_{m}$ & Number of APs in the $m$th subnetwork \\ \hline
         $\mathbf{g}_{k, \mathcal{B}_{m}}$ & Channel gain vector between user $u_{k}$ and the AP set $\mathcal{B}_{m}$ \\ \hline
         $\mathbf{w}_{k, \mathcal{B}_{m}}$ & Precoding vector between user $u_{k}$ and the AP set $\mathcal{B}_{m}$ \\ \hline
         $d_{k,l}$   & Euclidean distance between user $u_{k}$ and AP $b_{l}$ \\ \hline
         $N_{0}$       & Variance of additive white Gaussian noise (AWGN) \\ \hline
         $\alpha$        & Path-loss exponent \\ \hline
         $P$         & Average transmission power of each AP \\ \hline
         $\bar{R}$    & Average per-user rate \\ \hline
         $\bar{R}_{min}$    & Average minimum per-user rate \\ \hline
         $\bar{\eta} _{EE}$     & Average energy efficiency \\ \hline
         $\lambda^{*}$         & Optimal AP-selection ratio \\ \hline
     \end{tabular}}
 \end{table}

\section{System Model}
\vspace{2mm}
Consider the downlink transmission of a large-scale wireless network with a circular area of radius $D$, where a large amount of single-antenna APs and single-antenna users are randomly distributed. The set of APs is denoted by $\mathcal{B}=\left\{b_{1}, b_{2}, \cdots, b_{L}\right\}$ and the set of users is denoted by $\mathcal{U}=\left\{u_{1}, u_{2}, \cdots, u_{K}\right\}$ with $\left|\mathcal{B}\right|={L}$ and $\left|\mathcal{U}\right|={K}$, respectively. Table I summarizes the main notations used throughout this paper for ease of reading.

\subsection{Clustered Cell-Free Network Model}
\vspace{2mm}

Under the clustered cell-free network architecture, as illustrated in Fig. 1, a set of nonoverlapping subnetworks operating in parallel are formed by grouping APs and users, in each of which the APs jointly serve users to cancel out the intra-subnetwork interference. Note that there is no need to involve all the APs for signal transmission, as the APs far away from users contribute little to the desired signal power yet consume loads of energy and interfere with other subnetworks significantly. In this paper, we focus on improving the system energy efficiency of a clustered cell-free network with a fixed number of subnetworks by properly selecting APs for transmission, while the number of subnetworks should be determined according to practical requirements as detailed in \cite{JunyuanTWC, 8007415, 2022}. 
Let us define \emph{subnetwork} below first.

\begin{figure}[t]
	\centering                                    
	\includegraphics[width=0.48\textwidth]{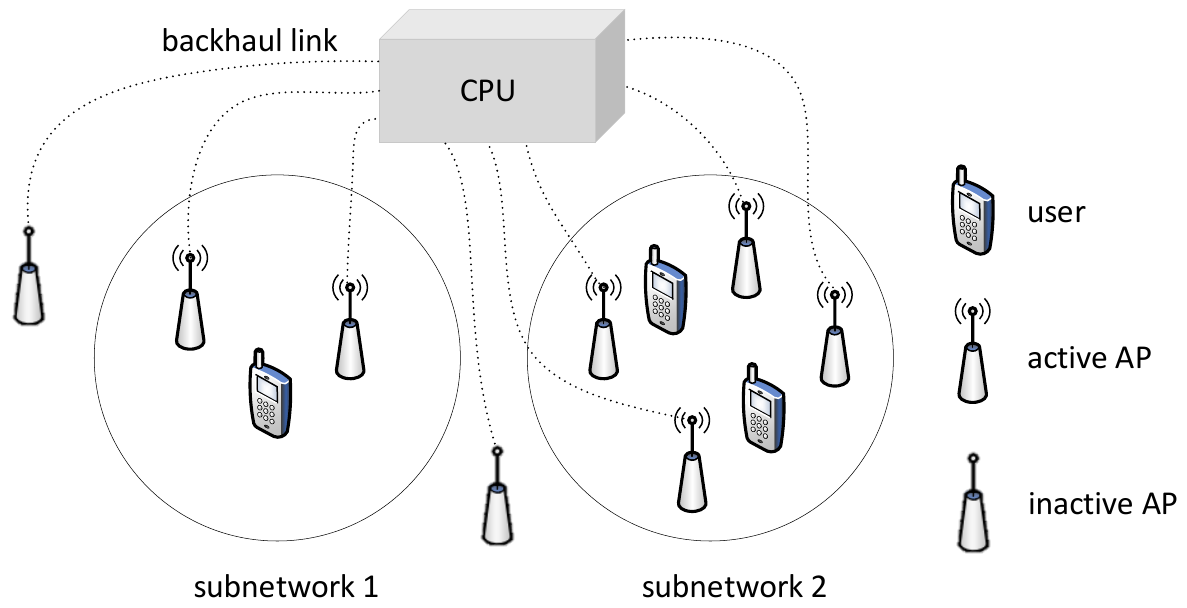}
	\caption{Graphical illustration of clustered cell-free network with AP selection.}
\end{figure}

\vspace{3mm}

\emph{Definition 1 (Subnetwork):
Denote a set of APs as $\mathcal{B}$, a set of users as $\mathcal{U}$, and the number of subnetworks as $M$. For each $m=1, 2, \cdots, M$, let the corresponding AP cluster be $\mathcal{B}_{m}\subseteq\mathcal{B}$ and the user cluster be $\mathcal{U}_{m}\subseteq\mathcal{U}$ with $\left|\mathcal{B}_{m}\right|={L}_{m}$ and $\left|\mathcal{U}_{m}\right|={K}_{m}$, respectively.
The $m$th subnetwork is then defined as $\mathcal{S}_{m}=\mathcal{B}_{m}\cup\hspace{1mm}\mathcal{U}_{m}$, if and only if $\mathcal{S}_{m}\cap\mathcal{S}_{n}=\emptyset$,
$\forall n\neq m$, $m=1, 2, \cdots, M$.}

Consider user $u_{k}$ in the $m$th subnetwork as the reference user.
The received signal at
user $u_{k}\in\mathcal{U}_{m}$ is given by
\begin{align}\label{2.1.1}
y_{k}=&\underbrace{\mathbf{g}_{k, \mathcal{B}_{m}} \mathbf{x}_{k, \mathcal{B}_{m}}}_{\text {desired signal }}
\hspace{1mm}+\underbrace{\sum_{u_{j} \in \mathcal{U}_{m}, j \neq k} \mathbf{g}_{k, \mathcal{B}_{m}} \mathbf{x}_{j, \mathcal{B}_{m}}}_{\text {intra-subnetwork interference }}\notag \\
&+\underbrace{\sum_{n =1, n \neq m}^{M}\sum_{u_{j} \in \mathcal{U}_{n}} \mathbf{g}_{k, \mathcal{B}_{n}} \mathbf{x}_{j, \mathcal{B}_{n}}}_{\text {inter-subnetwork interference }}+\hspace{1mm}z_{k},
\end{align}
where $\mathbf{x}_{k, \mathcal{B}_{m}}\in \mathbb{C}^{L_{m} \times 1}$
denotes the transmitted signal from the APs in the $m$th subnetwork to user $u_{k}$.
$z_{k} \sim \mathcal{C N}\left(0, N_{0}\right)$ represents the additive white Gaussian noise (AWGN) with zero mean and variance $N_{0}$.
$\mathbf{g}_{k, \mathcal{B}_{m}} \in \mathbb{C}^{1 \times L_{m}}$
is the channel gain vector between user $u_{k}$ and the AP set $\mathcal{B}_{m}$, which can be written as
\begin{equation}\label{2.1.2}
\mathbf{g}_{k, \mathcal{B}_{m}}=\boldsymbol{\gamma}_{k, \mathcal{B}_{m}} \circ \hspace{1mm}\mathbf{h}_{k, \mathcal{B}_{m}},
\end{equation}
where $\circ$ is the Hadamard product.
$\mathbf{h}_{k, \mathcal{B}_{m}} \in \mathbb{C}^{1 \times L_{m}}$
denotes the corresponding small-scale fading vector with entries modeled as independent and identically distributed (i.i.d) 
complex Gaussian random variables with zero mean and unit variance.
$\boldsymbol{\gamma}_{k, \mathcal{B}_{m}} \in \mathbb{R}^{1 \times L_{m}}$
represents the large-scale fading vector between user $u_{k}$ and the AP set $\mathcal{B}_{m}$. The large-scale fading coefficient between user $u_{k}$ and AP $b_{l}$, $\gamma_{k, l}$, can be modeled as 
\begin{equation}\label{2.1.3}
\gamma_{k, l}={d}_{k,l}^{-\frac{\alpha}{2}},
\end{equation}
where
\begin{equation}
{d_{k,l}}=\left\|\mathbf{r}_{l}^{B}-\mathbf{r}_{k}^{U}\right\|
\end{equation}
is the distance between user $u_{k}$ located at $\mathbf{r}_{k}^{U}$ and AP $b_{l}$ located at $\mathbf{r}_{l}^{B}$.
$\alpha$ denotes the path-loss exponent.

Assume that full CSI is perfectly known at both the transmitters and receivers,
and ZFBF is performed in each subnetwork for intra-subnetwork interference cancellation.
In the $m$th subnetwork,
the transmitted signal from the APs in $\mathcal{B}_{m}$ to user $u_{k}\in\mathcal{U}_{m}$, 
$\mathbf{x}_{k, \mathcal{B}_{m}}$, is given by
\begin{equation}\label{2.2.1}
\mathbf{x}_{k, \mathcal{B}_{m}}=\mathbf{w}_{k, \mathcal{B}_{m}} \cdot\hspace{1mm} s_{k},
\end{equation}
where $s_{k}$ represents the information-bearing signal for user $u_{k}$ with $\mathbb{E}[s_{k}s_{k}^{*}]=P_{k}$, and
$\mathbf{w}_{k, \mathcal{B}_{m}}$ denotes the precoding vector with $\left\|\mathbf{w}_{k, \mathcal{B}_{m}}\right\|=1$.
The total transmission power in each subnetwork $\mathcal{S}_{m}$ is assumed to be fixed at $P{L}_{m}$, which is proportional to the number of APs $L_{m}$ in it with $P$ denoting the average transmission power per AP.
With equal power allocation over users in each subnetwork,
the transmission power $P_{k}$ for user $u_{k}\in\mathcal{U}_{m}$ can be obtained as
\begin{equation}\label{2.2.2}
P_{k}=\frac{P{L}_{m}}{{K}_{m}}.
\end{equation}
By denoting the channel gain matrix between user cluster $\mathcal{U}_{m}$ 
and its serving AP set $\mathcal{B}_{m}$ as $\mathbf{G}_{\mathcal{U}_{m}} \in \mathbb{C}^{{K}_{m} \times{L}_{m}}$,
the pseudo-inverse of $\mathbf{G}_{\mathcal{U}_{m}}$, $\mathbf{F}_{\mathcal{U}_{m}}$, is given by
\begin{equation}\label{2.2.3}
\mathbf{F}_{\mathcal{U}_{m}}=\mathbf{G}_{\mathcal{U}_{m}}^{\dagger}\left(\mathbf{G}_{\mathcal{U}_{m}} \mathbf{G}_{\mathcal{U}_{m}}^{\dagger}\right)^{-1}.
\end{equation}
As the number of APs $L_{m}$ is required to be no smaller than the number of users $K_{m}$ in each subnetwork to apply ZFBF,
the corresponding precoding vector $\mathbf{w}_{k, \mathcal{B}_{m}}$ is then\cite{7031453}
\begin{equation}\label{2.2.4}
\mathbf{w}_{k, \mathcal{B}_{m}}=\left\{\begin{array}{ll}
\frac{\mathbf{f}_{k, \mathcal{U}_{m}}}{\left\|\mathbf{f}_{k, \mathcal{U}_{m}}\right\|} & \text { if }{L}_{m}\geq{K}_{m}, \\
\mathbf{0}_{{L}_{m} \times 1} & \text{ otherwise},
\end{array}\right.
\end{equation}
where $\mathbf{f}_{k, \mathcal{U}_{m}}$ is the column vector of $\mathbf{F}_{\mathcal{U}_{m}}$ with respect to user $u_{k}$.
Since the intra-subnetwork interference can be canceled by adopting ZFBF,
there is only inter-subnetwork interference in the system.
With a large amount of APs and users, the power of the inter-subnetwork interference received at user $u_{k}\in\mathcal{U}_{m}$, $I_{k}$, can be obtained from (\ref{2.1.1}) and (\ref{2.2.1}) as
\begin{equation}\label{2.2.5}
I_{k}=\sum_{n =1, n \neq m}^{M}\sum_{u_{j} \in \mathcal{U}_{n}} \mathbf{g}_{k, \mathcal{B}_{n}} \mathbf{w}_{j, \mathcal{B}_{n}} \mathbf{w}_{j, \mathcal{B}_{n}}^{\dagger} \mathbf{g}_{k, \mathcal{B}_{n}}^{\dagger} P_{j}.
\end{equation}

\subsection{Energy Efficiency}
By normalizing the total system bandwidth into unity,
the achievable ergodic rate of reference user $u_{k} \in \mathcal{U}_{m}$ can be obtained as
\begin{equation}\label{2.2.6}
R_{k}=\mathbb{E}_{\mathbf{H}}\left[\log_{2}\left(1+\frac{\mathbf{g}_{k, \mathcal{B}_{m}} \mathbf{w}_{k, \mathcal{B}_{m}} \mathbf{w}_{k, \mathcal{B}_{m}}^{\dagger} \mathbf{g}_{k, \mathcal{B}_{m}}^{\dagger}P_{k}}{N_{0}+I_{k}}\right)\right],
\end{equation}
where $\mathbf{H}\in \mathbb{C}^{K \times L}$ 
denotes the small-scale fading matrix between all the $L$ APs and $K$ users.
The sum rate of users, $R_{sum}$, can be written as
\begin{equation}\label{2.2.7}
R_{sum}=\sum_{k=1}^{K} R_{k},
\end{equation}
and the energy efficiency $\eta_{EE}$ of the system is given by
\begin{equation}\label{2.3.5}
\eta_{EE}=\frac{R_{sum}}{P_{T}},
\end{equation}
where $P_{T}$ is the total power consumption for the data transmission,
consisting of the total power consumed by the APs, $P_{A}$,
and the backhaul power consumption $P_{B}$\cite{6502484}, i.e.,
\begin{equation}\label{2.3.1}
P_{T}=P_{A}+P_{B}.
\end{equation}
The power consumption of APs in all the $M$ subnetworks can be obtained as \cite{7062017}
\begin{equation}\label{2.3.2}
P_{A}=\sum_{m=1}^{M}\left(\frac{PL_{m}}{\tau}+P_{c}L_{m}\right),
\end{equation}
where $\tau$ represents the power amplifier efficiency, and $P_{c}$ is the circuit power consumption incurred by frequency synthesizers, mixers and transmit filters, etc.
The power consumed in
the backhaul link, $P_{B}$, can be written as \cite{9148948}
\begin{equation}\label{2.3.3}
P_{B}=P_{\emph{fix}}L+P_{b} R_{sum},
\end{equation}
where $P_{\emph{fix}}$ is the power consumed regardless of the backhaul traffic, and $P_{b}$ denotes the
traffic-dependent power consumption in the backhaul.
By substituting (\ref{2.3.2}) and (\ref{2.3.3}) into (\ref{2.3.1}), 
the total power consumption $P_{T}$ can be obtained as
\begin{equation}\label{2.3.4}
P_{T}=\left(\frac{P}{\tau}+P_{c}\right)\sum_{m=1}^{M}L_{m}+P_{\emph{fix}}L+P_{b} R_{sum}.
\end{equation}

Since the APs and users are randomly distributed, the sum rate $R_{sum}$ varies with the layout of APs and users, and thus the energy efficiency fluctuates.
To see how the system parameters affect the energy efficiency, in this paper, we focus on the average energy efficiency over the locations of APs and users, $\bar{\eta} _{EE}$, defined as
\begin{equation}\label{2.3.6}
\bar{\eta} _{EE}\triangleq \mathbb{E}_{{\left\{\mathbf{r}_{k}^{U}\right\}_{u_{k} \in \mathcal{U}}}, {\left\{\mathbf{r}_{l}^{B}\right\}_{b_{l} \in \mathcal{B}}}}\left[\eta_{EE}\right].
\end{equation}

\section{Clustering for Cell-Free Networking with AP Selection}
In this section, a novel user-centric ratio-fixed AP-selection
based clustering algorithm will be proposed to incorporate AP selection into clustered
cell-free networking for energy saving. To figure out a simple yet effective clustering strategy for clustered cell-free networking, let us start by assuming that all the APs are involved for signal transmission.

\subsection{AP-Centric or User-Centric?}
The subnetworks can be formed by partitioning APs and users, which is usually performed in an AP-centric or user-centric manner as detailed in the following, where agglomerative hierarchical clustering is adopted to group APs or users.$\footnote{
Note that agglomerative hierarchical clustering is adopted here as it achieves the best performance among popular clustering algorithms. Detailed discussions on the effects of clustering algorithms
are presented in Section III.-B.}$

\begin{itemize}
\item \emph{AP-centric clustering}: The APs are first clustered into a set of groups, 
and then each user is affiliated with its best AP with the highest
large-scale fading coefficient to form subnetworks.
\item \emph{User-centric clustering}: The users are first clustered into a set of groups,
and then each AP is affiliated with its best user with the highest
large-scale fading coefficient to form subnetworks.
\end{itemize}
\begin{table}[t]
\centering 
\caption{Simulation Parameters}
\begin{tabular}{|l|l|}
\hline
Cell radius $D$                               & 1000m    \\ \hline
Noise power $N_{0}$                    & -104dBm  \\ \hline
Path-loss exponent $\alpha$        & 4              \\ \hline
Transmission power per-AP $P$        & 2W       \\ \hline
Circuit power consumption $P_{c}$             & 1W       \\ \hline
Fixed power consumption $P_{\emph{fix}}$             & 0.05W    \\ \hline
Traffic-dependent power consumption $P_{b}$ & 0.1W/bps/Hz \\ \hline
Power amplifier efficiency $\tau$             & 38$\%$      \\ \hline
\end{tabular}
\end{table}
\begin{figure}[t]
\centering                                    
\includegraphics[width=0.45\textwidth]{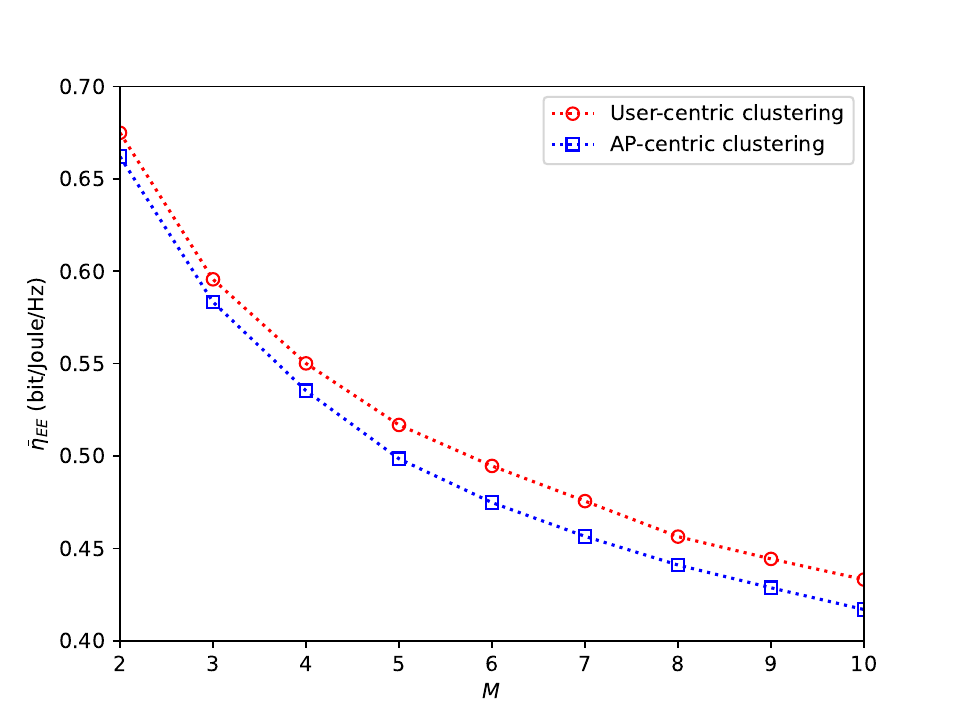}\\
\caption{Average energy efficiency $\bar{\eta} _{EE}$ versus the number of subnetworks $M$ with AP-centric and user-centric clustering algorithms averaged over 1000 random realizations of AP and user positions. $L=200$. $K=100$.}
\end{figure}

Fig. 2 presents the average energy efficiency $\bar{\eta} _{EE}$ versus the number of subnetworks $M$
with both the AP-centric and user-centric clustering algorithms. The simulation parameters are summarized in Table II, which are used in all the simulations presented in this paper. Note that the typical values of power
parameters are taken from \cite{9148948} and \cite{7062017}.
As we can see from Fig. 2, the average energy efficiency $\bar{\eta} _{EE}$ with both the AP-centric and user-centric clustering algorithms decreases
as the number of subnetworks $M$ increases due to the increased inter-subnetwork interference. Moreover, the user-centric clustering algorithm
achieves higher average energy efficiency than the AP-centric approach. This is because with user-centric clustering, users strongly interfering with each other are always grouped into the same
subnetwork and each user is surrounded by some APs within the subnetwork. As a result, there are no users located at the subnetwork edge and thus can avoid the subnetwork-edge problem. By contrast,
with AP-centric clustering, the subnetwork-edge problem inherent to the cellular architecture still remains, leading to lower average energy efficiency.

\subsection{Effect of Clustering Algorithm}

Apart from the agglomerative hierarchical clustering algorithm adopted in the last section, which first treats each object as a singleton group and then successively merges two groups that are the most similar until it meets the termination condition, there are various other popular clustering algorithms that can be employed for clustered cell-free networking, such as
K-means and spectral clustering.\footnote{K-means assigns data points to clusters by minimizing the within-cluster sum of squared distances \cite{4767478}, and spectral clustering groups data based on the eigenvalue decomposition of Laplacian matrix derived from the data set \cite{von2007tutorial}.} Since users' positions are usually not available in practical mobile
communication systems, in this paper, the large-scale fading vectors are used for clustering. Specifically, let $\boldsymbol{\gamma}_{k}$ (dB) $\in \mathbb{R}^{1 \times L}$ denote the large-scale fading vector from all $L$ APs to user $u_{k}$ on a dB scale, the distance $s_{k, j}$ between the large-scale fading vectors $\boldsymbol{\gamma}_{k}$ and $\boldsymbol{\gamma}_{j}$ of users $u_{k}$ and $u_{j}$ can be measured by cosine distance, which is defined as \cite{7915506}:
\begin{equation}
s_{k, j}=1-\frac{\boldsymbol{\gamma}_{k}\cdot\boldsymbol{\gamma}_{j}^{T}}{\left\|\boldsymbol{\gamma}_{k}\right\|\cdot\|\boldsymbol{\gamma}_{j}\|}.
\end{equation}

To figure out which one works the best, 
Fig. 3 presents the average energy efficiency $\bar{\eta}
 _{EE}$ achieved with K-means, spectral clustering and agglomerative hierarchical clustering, when user-centric clustering is adopted due to its superior performance.
It is shown in Fig. 3 that agglomerative hierarchical clustering achieves the highest average energy efficiency. The reason is that compared to K-means, agglomerative hierarchical clustering does not rely on the selection of the initial center points and thus can avoid falling into local optima, while spectral clustering performs worse as it may lead to information loss due to dimensionality reduction.

\begin{figure}[t]
	\centering                                    
	\includegraphics[width=0.45\textwidth]{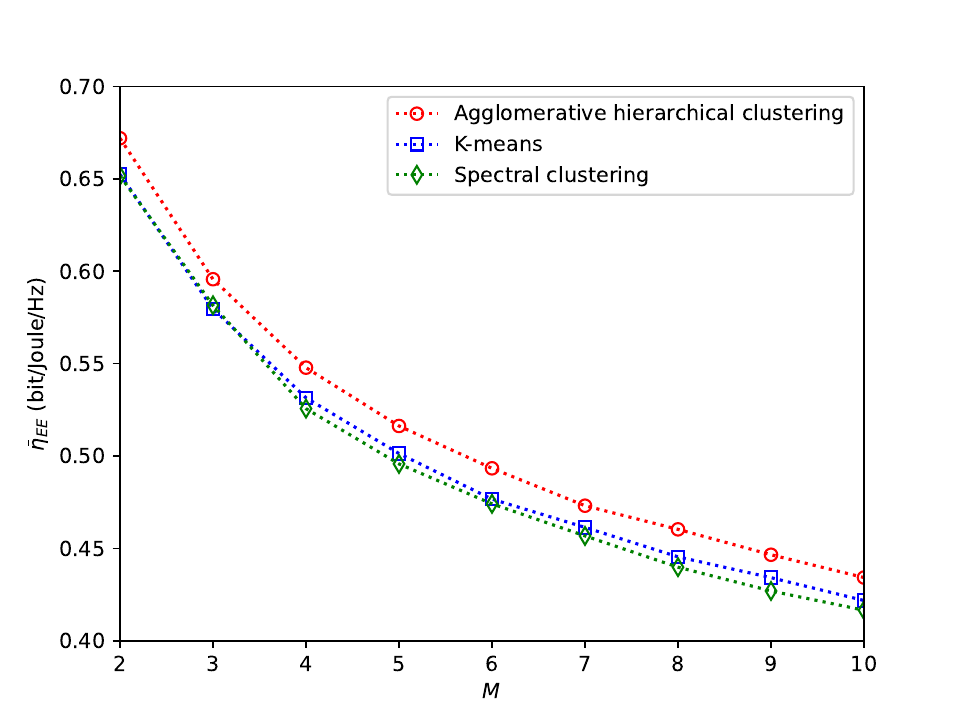}\\
	\caption{Average energy efficiency $\bar{\eta} _{EE}$ versus the number of subnetworks $M$ with popular clustering algorithms, including K-means, spectral clustering and agglomerative hierarchical clustering. $L=200$. $K=100$.}
\end{figure}

\subsection{User-Centric Ratio-Fixed AP-Selection Based Clustering}
It can be concluded from Figs. 2 and 3 that the user-centric clustering approach which adopts agglomerative hierarchical clustering achieves the highest average energy efficiency and hence should be adopted for clustered cell-free networking. 

To further facilitate adaptive AP selection, we propose to select $\lambda K$ out of $L$ APs, where $\lambda$ is defined as the AP-selection ratio. A simple way is to select $\lambda K$ APs with the highest large-scale fading coefficients after clustering users and assign each of them to the cluster to which its best user with the highest large-scale fading coefficient belongs. 
This algorithm is referred to as user-centric clustering with AP selection (UC-ApSel) in this paper. However, such a simple AP selection strategy usually leads to unfair services among users and even service outage, as most active APs could be assigned to a small number of subnetworks.

\begin{algorithm}[t]
	\caption{User-Centric Ratio-Fixed AP-Selection Based Clustering (UCR-ApSel)} 
	\begin{algorithmic}[1] 
		\REQUIRE AP set $\mathcal{B}=\left\{b_{1}, b_{2}, \cdots, b_{L}\right\}$, user set $\mathcal{U}=\left\{u_{1}, u_{2}, \cdots, u_{K}\right\}$, the number of subnetworks $M$, AP-selection ratio $\lambda$; \\
		\hspace*{-0.23in}{\bf Initialization:} Set each user as a cluster, i.e., $\mathcal{U}_{m}=\left\{u_{k}\right\}$, $\forall m=k$, $k=1, 2, \cdots, K$, $C=K$, $N=0$;
		\WHILE{$C>M$}
		\STATE {Calculate the cosine distance between any two clusters with average linkage by $\frac{\sum_{u_{k} \in \mathcal{U}_{m}}\sum_{u_{j} \in \mathcal{U}_{n}}s_{k, j}}{\left|\mathcal{U}_{m}\right|\cdot\left|\mathcal{U}_{n}\right|}$};    	
		\STATE {Merge the two user clusters with the minimum distance, $C=C-1$};
		\STATE {Rename the user clusters as $\mathcal{U}_{1}, \mathcal{U}_{2}, \cdots, \mathcal{U}_{C}$};
		\ENDWHILE
		\FOR{$m=1$ to $M$} 
		\STATE {$\mathcal{S}_{m}=\mathcal{U}_{m}$, ${K}_{m}=\left|\mathcal{U}_{m}\right|$, $N=N+\lfloor{K}_{m}\times \lambda \rfloor$};
		\ENDFOR 
		\WHILE{$\sum_{m =1}^{M}\left|\mathcal{S}_{m}\right|\textless K+N$}
		\STATE {$\left(b_{{l}^{*}}, u_{{k}^{*}}\right)=\arg \mathop{\max}_{b_{l}\in \mathcal{B}, u_{k}\in \mathcal{U}}\gamma_{k, l}$};
		\STATE {Denote $\mathcal{U}_{m^{*}}$ as the user cluster that $u_{k^{*}}$ belongs to};
		\IF{$\left|\mathcal{S}_{m^{*}}\right|\textless \lfloor{K}_{m^{*}}\times(1+\lambda)\rfloor$} 
		\STATE {$\mathcal{S}_{m^{*}}=\mathcal{S}_{m^{*}}\cup\hspace{1mm}b_{l^{*}}$, $\mathcal{B}=\mathcal{B}\hspace{1mm}\backslash\hspace{1mm}b_{l^{*}}$};
		\ELSE 
		\STATE {$\mathcal{U}=\mathcal{U}\hspace{1mm}\backslash\hspace{1mm}\mathcal{U}_{m^{*}}$};
		\ENDIF 
		\ENDWHILE
		\ENSURE $\left\{\mathcal{S}_{1}, \mathcal{S}_{2}, \cdots, \mathcal{S}_{M}\right\}$
	\end{algorithmic} 
\end{algorithm}

\begin{figure}[t]
	\centering                                    
	\includegraphics[width=0.45\textwidth]{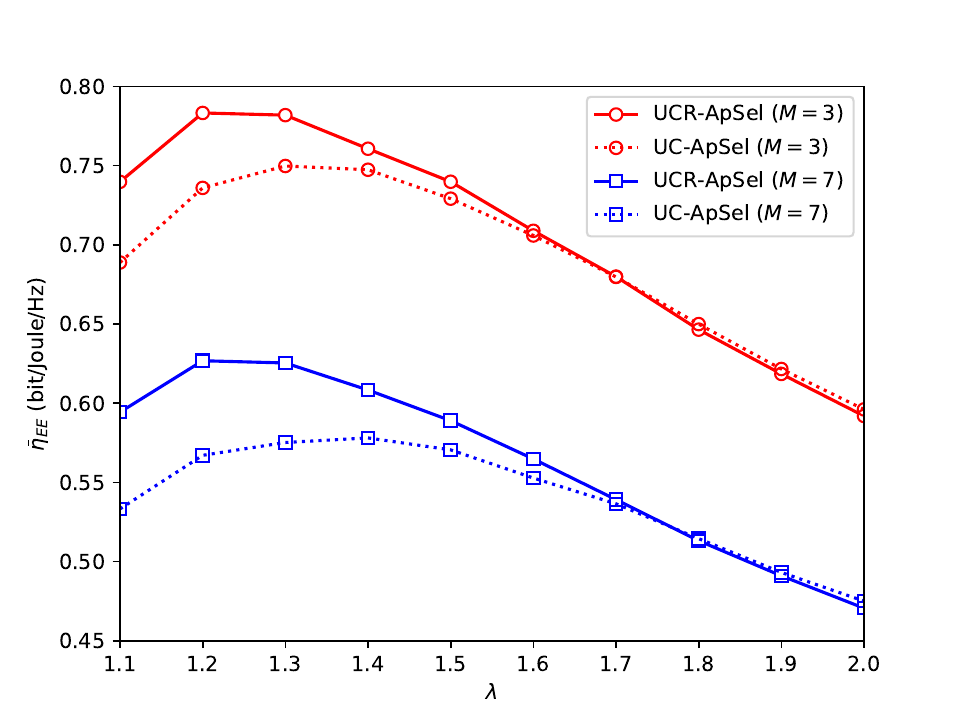}\\
	\caption{Average energy efficiency $\bar{\eta} _{EE}$ versus the AP-selection ratio $\lambda$ with both the proposed UCR-ApSel algorithm and the benchmarking UC-ApSel algorithm. $L=200$. $K=100$.}
\end{figure}

In order to achieve uniform performance across the users in different subnetworks, we propose to fix the ratio of the number of APs to the number of users in each subnetwork at some constant $1<\lambda\leq L/K$.\footnote{
Note that the AP-selection ratio $\lambda>1$ is assumed in this paper to guarantee that ZFBF can always be applied to cancel out the intra-subnetwork interference successfully.} A two-step user-centric ratio-fixed AP-selection based clustering (UCR-ApSel) algorithm is then proposed. In the first step, the users in the system are grouped into $M$ clusters by adopting the agglomerative hierarchical clustering algorithm with cosine distance. In the second step, APs are selected and assigned to user clusters in a greedy manner. Specifically, the AP-user pair $\left(b_{{l}^{*}}, u_{{k}^{*}}\right)$ with the highest large-scale fading coefficient between users and available APs, i.e., $\left(b_{l^{*}}, u_{k^{*}}\right)=\arg \mathop{\max}\gamma_{k, l}$ is searched and AP $b_{{l}^{*}}$ is assigned to the cluster containing user $u_{{k}^{*}}$ in each iteration until the ratio of the number of APs to the number of users increases to $\lambda$ in each subnetwork.
The complete process of our proposed UCR-ApSel algorithm is summarized in Algorithm 1.

\begin{figure*}[t]
	\centering
	\subfloat[]{
		\label{1a}
		\includegraphics[width=0.45\textwidth]{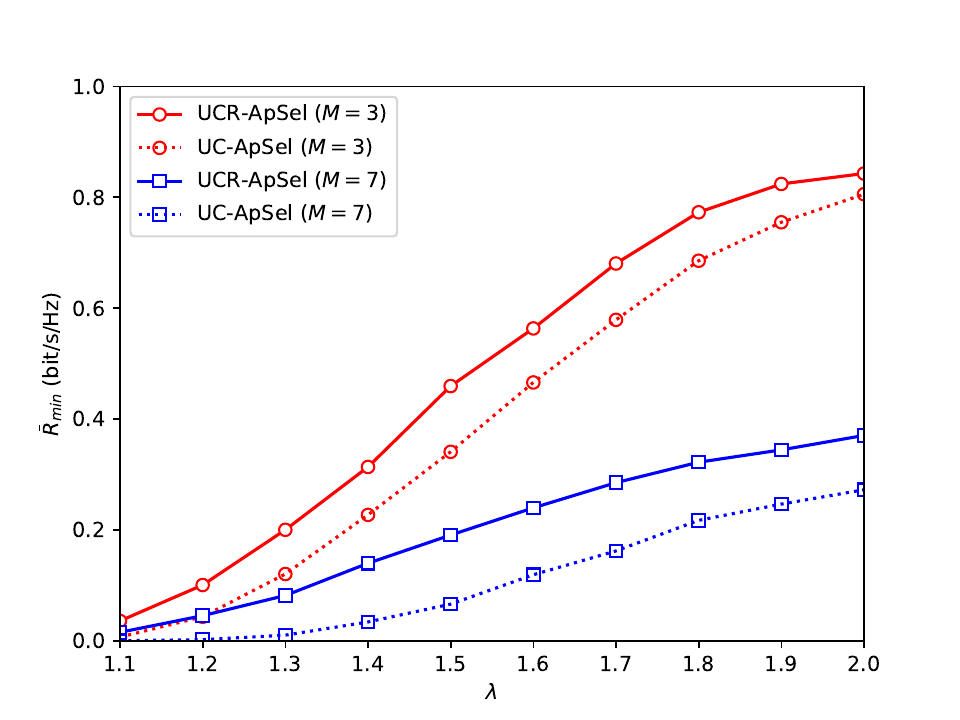}}  
  \hspace{8mm}
	\subfloat[]{
		\label{1b}
		\includegraphics[width=0.45\textwidth]{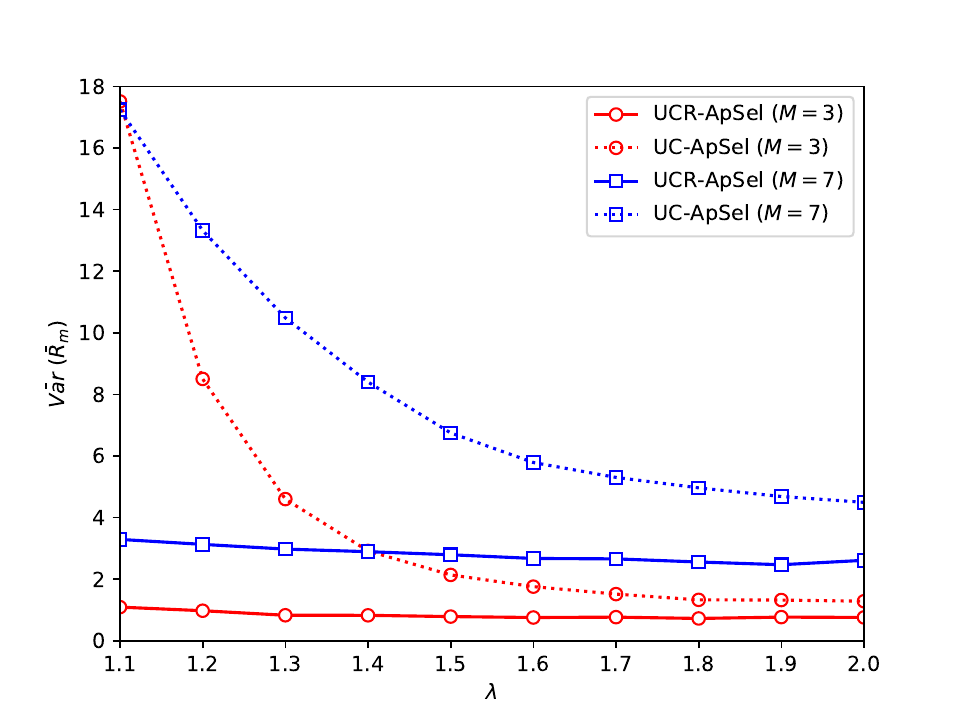}}
	\caption{(a) Average minimum rate of users $\bar{R}_{min}$ and (b) average variance of the per-user rate in different subnetworks $\bar{V\hspace{-0.75mm}ar}\left(\bar{R}_{m}\right)$ versus AP-selection ratio $\lambda$ with both the proposed UCR-ApSel algorithm and the benchmarking UC-ApSel algorithm. $L=200$. $K=100$.}
\end{figure*}

Fig. 4 depicts the average energy efficiency $\bar{\eta} _{EE}$ versus AP-selection ratio $\lambda$ with both the proposed UCR-ApSel algorithm and the benchmarking UC-ApSel algorithm. We can note that as the AP-selection ratio $\lambda$ increases, the average energy efficiency $\bar{\eta} _{EE}$ increases firstly  and then decreases, corroborating that there is no need to involve the APs far away from users for signal transmission. It is also shown that the proposed UCR-ApSel algorithm achieves higher average energy efficiency than the UC-ApSel algorithm in the range of $\lambda\leq1.7$.
As can be seen in Fig. 4, the average energy efficiency $\bar{\eta} _{EE}$ with the proposed UCR-ApSel algorithm is $11\%$ higher than that with UC-ApSel when the AP-selection ratio $\lambda=1.1$ and the number of subnetworks $M=7$.
However, the benchmarking UC-ApSel algorithm achieves slightly higher energy efficiency when $\lambda>1.7$. This is due to the fact that with a large $\lambda$, certain APs might be assigned to the user clusters considerably far from them with the proposed UCR-ApSel algorithm as the ratio of the number of APs to the number of users is forced to be the same across subnetworks, causing energy efficiency degradation.

To have a look at the effectiveness of fixing the ratio of the number of APs to the number of users across different subnetworks in improving user fairness, Fig. 5 presents the average minimum rate of users $\bar{R}_{min}\hspace{-1mm}\triangleq\hspace{-0.5mm}\mathbb{E}_{{\hspace{-0.5mm}\left\{\hspace{-0.5mm}\mathbf{r}_{k}^{U}\hspace{-0.5mm}\right\}_{\hspace{-0.5mm}u_{k} \in \mathcal{U}}}, {\hspace{-0.5mm}\left\{\hspace{-0.5mm}\mathbf{r}_{l}^{B}\hspace{-0.5mm}\right\}_{\hspace{-0.5mm}b_{l} \in \mathcal{B}}}}$
$\left[\min_{u_{k} \in \mathcal{U}}R_{k}\right]$ and the average variance of the per-user rate in different subnetworks $\bar{V\hspace{-0.75mm}ar}\left(\bar{R}_{m}\right)\triangleq\hspace{-0.5mm}\mathbb{E}_{{\hspace{-0.5mm}\left\{\hspace{-0.5mm}\mathbf{r}_{k}^{U}\hspace{-0.5mm}\right\}_{\hspace{-0.5mm}u_{k} \in \mathcal{U}}}, {\hspace{-0.5mm}\left\{\hspace{-0.5mm}\mathbf{r}_{l}^{B}\hspace{-0.5mm}\right\}_{\hspace{-0.5mm}b_{l} \in \mathcal{B}}}}$
$V\hspace{-0.75mm}ar\left\{\sum_{k=1}^{K_{m}}R_{k}/K_{m}\right\}_{m=1, 2, \cdots, M}$ versus AP-selection ratio $\lambda$ with both the proposed UCR-ApSel algorithm and the benchmarking UC-ApSel algorithm.
We can see from Fig. 5(a) that our proposed UCR-ApSel algorithm improves the 
average minimum rate $\bar{R}_{min}$ significantly compared to the UC-ApSel approach. In addition, it can be observed from Fig. 5(b) that the average variance of the per-user rate performance in different subnetworks with the proposed UCR-ApSel algorithm is much lower than that with the UC-ApSel algorithm, especially when the AP-selection ratio $\lambda$ is small. The comparison implies that balanced services among subnetworks and uniform rate performance among users can be achieved by applying the
proposed UCR-ApSel algorithm thanks to the balanced AP assignment.

It can be concluded from the above discussion that by fixing the ratio of the number of APs to the number of users in each subnetwork as a constant,
the proposed UCR-ApSel algorithm for
clustered cell-free networking improves both energy efficiency and user fairness. More importantly, it is shown in Fig. 4 that there exists an optimal AP-selection ratio that maximizes the average energy efficiency. How to obtain the optimal AP-selection ratio for energy-efficient clustered cell-free networking is of great practical importance, which will be studied carefully in Section IV.

\subsection{Complexity Analysis}
The proposed UCR-ApSel algorithm consists of two steps: 1) user clustering and 2) AP selection. In the first step, $K$ users are clustered into $M$ subnetworks by adopting agglomerative hierarchical clustering. The corresponding computational complexity is $O(K^{3})$ \cite{7583663}. In the second step of AP selection, $\lambda K$ APs are selected and assigned to user clusters in a greedy manner. In each iteration, the best AP-user pair with the highest large-scale fading coefficient is first selected and then the AP is assigned to the subnetwork containing its paired user if the ratio of the number of APs to the number of users is smaller than a given threshold $\lambda$. To find out the best AP-user pair, at most $LK-1$ comparisons are required. To assign $\lambda K$ APs to $M$ subnetworks, at most $\lambda K+M-1$ iterations are needed. Since the number of subnetworks $M$ is usually much smaller than the number of users $K$, i.e., $M\ll K$, the complexity of AP selection is $O((LK-1)(\lambda K+M-1))=O(\lambda LK^{2})$. The total computational complexity of the proposed UCR-ApSel algorithm is $O(K^{3}+\lambda LK^{2})=O(\lambda LK^{2})$, as the total number of APs $L$ is much larger than the number of users $K$.

\section{Average Energy Efficiency Analysis}
In order to derive an explicit expression of the optimal AP-selection ratio $\lambda^{*}$ that maximizes the average energy efficiency of the clustered cell-free network, it is essential to study the average energy efficiency $\bar{\eta}_{EE}$ achieved with the proposed UCR-ApSel algorithm.
It is clear from (\ref{2.2.6})-(\ref{2.3.5}) that the energy efficiency $\eta_{EE}$ is determined by the  large-scale fading coefficients of both signal and interference channels. Since the subnetworks in clustered cell-free networks are formed by
dynamically clustering APs and users, the network structure varies with AP and user positions,
making it extremely difficult and even intractable to analyze the average energy efficiency. Therefore, in this paper, we resort to an upper-bound to investigate how the AP-selection ratio $\lambda$ affects the average energy efficiency $\bar{\eta}_{EE}$.

\subsection{A Closed-Form Upper-Bound of Average Energy Efficiency $\bar{\eta}_{EE}^{ub}$}
As $f\left(x\right)=\frac{x}{a+bx}$ with $x>0$ is a concave function of $x$ for $a>0$ and $b>0$,
according to Jensen’s inequality, the average energy efficiency $\bar{\eta}_{EE}$ can be obtained by combining (\ref{2.2.7}),
(\ref{2.3.5}) and (\ref{2.3.6}) as
\begin{equation}\label{4.2.3}
\bar{\eta}_{EE}\leq\frac{K\bar{R}}{\left(\frac{P}{\tau}+P_{c}\right)\sum_{m=1}^{M}L_{m}+P_{\emph{fix}} L+P_{b} K\bar{R}},
\end{equation}
where
\begin{equation}\label{4.2.4}
\bar{R}=\mathbb{E}_{{\left\{\mathbf{r}_{k}^{U}\right\}_{u_{k} \in \mathcal{U}}}, {\left\{\mathbf{r}_{l}^{B}\right\}_{b_{l} \in \mathcal{B}}}}
\left[R_{k}\right]
\end{equation}
is the average per-user rate.
\begin{figure}[!t]
	\centering                                
	\includegraphics[width=0.45\textwidth]{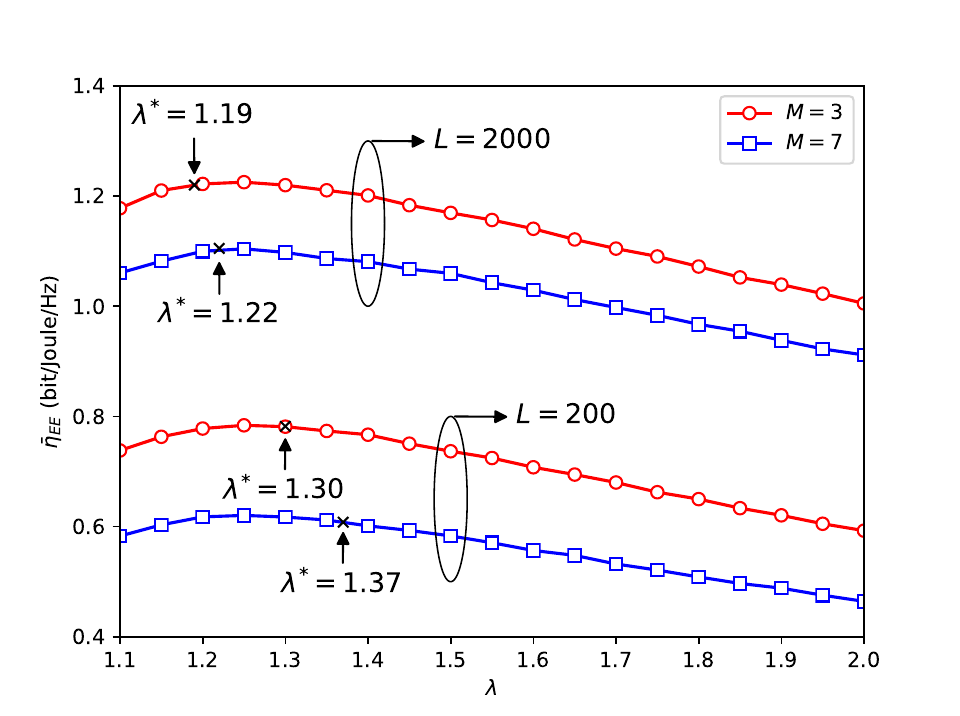}
	\caption{Average energy efficiency $\bar{\eta}_{EE}$ versus the AP-selection ratio $\lambda$. The optimal $\lambda^{*}$ given in (\ref{c.3}) is presented.
		$K=100$.}
\end{figure}

With the ratio of the number of APs to the number of users fixed at $\lambda$ in all the subnetworks, we have
\begin{equation}\label{4.2.11}
L_{m}/K_{m}=\lambda, 
\end{equation}
for any $m=1, 2, \cdots, M$. The average per-user rate $\bar{R}$ can be obtained by combining (\ref{2.2.2}), (\ref{2.2.5})-(\ref{2.2.6}) and (\ref{4.2.4})-(\ref{4.2.11}) as
\begin{equation}\label{4.1.1}
\bar{R}\stackrel{P/N_{0}\gg 1}{\approx}\bar{R}_{1}-\bar{R}_{2},
\end{equation}
where
\begin{equation}\label{4.1.2}
\bar{R}_{1}\hspace{-1mm}=\hspace{-0.5mm}\mathbb{E}_{\mathbf{H}, {{\left\{\hspace{-0.5mm}\mathbf{r}_{k}^{U}\hspace{-0.5mm}\right\}_{u_{k} \in \mathcal{U}}}, {\left\{\hspace{-0.5mm}\mathbf{r}_{l}^{B}\hspace{-0.5mm}\right\}_{b_{l} \in \mathcal{B}}}}}\hspace{-1mm}\left[\log_{2}\hspace{-1mm}\left(\hspace{-0.5mm}\mathbf{g}_{k, \mathcal{B}_{m}}\hspace{-0.5mm} \mathbf{w}_{k, \mathcal{B}_{m}}\hspace{-0.5mm}\mathbf{w}_{k, \mathcal{B}_{m}}^{\dagger}\hspace{-0.5mm} \mathbf{g}_{k, \mathcal{B}_{m}}^{\dagger}\hspace{-0.5mm}\right)\hspace{-1mm}\right],
\end{equation}
and
\begin{equation}\bar{R}_{2}\hspace{-1.25mm}=\hspace{-1mm}\mathbb{E}_{\hspace{-0.125mm}\mathbf{H}\hspace{-0.25mm}, \hspace{-0.5mm}{{\left\{\hspace{-0.5mm}\mathbf{r}_{k}^{U}\hspace{-0.75mm}\right\}_{\hspace{-0.75mm}u_{\hspace{-0.25mm}k}\hspace{-0.5mm} \in\hspace{-0.125mm} \mathcal{U}}}\hspace{-0.5mm},\hspace{-0.5mm} {\left\{\hspace{-0.5mm}\mathbf{r}_{l}^{B}\hspace{-0.75mm}\right\}_{\hspace{-0.75mm}b_{\hspace{-0.125mm}l}\hspace{-0.5mm} \in\hspace{-0.25mm} \mathcal{B}}}}}\hspace{-1.75mm}\left[\hspace{-0.75mm}\log _{\hspace{-0.25mm}2}\hspace{-1.5mm}\left(\hspace{-0.25mm}\sum_{\substack{n =1\\n \neq m}}^{M}\hspace{-0.75mm}\sum_{u_{j} \in \mathcal{U}_{n}}\hspace{-2mm}\mathbf{g}_{k\hspace{-0.25mm}, \mathcal{B}_{n}}\hspace{-0.75mm}\mathbf{w}_{j\hspace{-0.25mm}, \mathcal{B}_{n}}\hspace{-0.75mm}\mathbf{w}_{j\hspace{-0.25mm}, \mathcal{B}_{n}}^{\dagger}\hspace{-0.75mm}\mathbf{g}_{k\hspace{-0.25mm}, \mathcal{B}_{n}}^{\dagger}\hspace{-1.5mm}\right)\hspace{-1.25mm}\right].
\end{equation}
\begin{figure*}[!hb]
	\begin{normalsize} 
 		\hrulefill
		\begin{align}\label{4.2.2}
	\bar{\eta}_{EE}\leq\bar{\eta}_{EE}^{ub}=\frac{\log _{2}\left((\lambda-1)/\lambda+M/\left(\lambda K\right)\right)+\log _{2}\left(L/M\right)+\gamma\log _{2}{e}}{\frac{2}{\alpha}\left(\frac{P}{\tau}+P_{c}\right)\lambda+\frac{2}{\alpha}P_{\emph{fix}} L/K+P_{b}\left[\log _{2}\left((\lambda-1)/\lambda+M/\left(\lambda K\right)\right)+\log _{2}\left(L/M\right)+\gamma\log _{2}{e}\right]}\tag{29}
		\end{align}
  			\hrulefill
	\begin{align}\label{4.2.21}
		\bar{\eta}_{EE}^{ub}\stackrel{P\gg P_{fix}, K\gg M}{\approx}\frac{\log _{2}\left(1-1/\lambda\right)+\log _{2}\left(L/M\right)+\gamma\log _{2}{e}}{\frac{2}{\alpha}\left(\frac{P}{\tau}+P_{c}+P_{\emph{fix}}\right)\lambda+P_{b}\left(\log _{2}\left(1-1/\lambda\right)+\log _{2}\left(L/M\right)+\gamma\log _{2}{e}\right)}\tag{30}
	\end{align}
   		\hrulefill
    \begin{equation}\label{c}
	\frac{\partial \bar{\eta}_{EE}^{ub}}{\partial \lambda}=\frac{\frac{2}{\alpha}\left(\frac{P}{\tau}+P_{c}+P_{\emph{fix}}\right)\left(\log _{2}{e}/\left(\lambda-1\right)-\log _{2}\left(1-1/\lambda\right)-\log _{2}\left(L/M\right)-\gamma\log _{2}{e}\right)}{\left[\frac{2}{\alpha}\left(\frac{P}{\tau}+P_{c}+P_{\emph{fix}}\right)\lambda+P_{b}\left(\log _{2}\left(1-1/\lambda\right)+\log _{2}\left(L/M\right)+\gamma\log _{2}{e}\right)\right]^{2}}\tag{32}
\end{equation}
	\end{normalsize}
\end{figure*}

Appendix A shows that $\bar{R}_{1}$ can be approximated as
\begin{equation}\label{4.1.5}
\bar{R}_{1}\approx\mathbb{E}\left[\log _{2}\left({d}_{k, {l}^{*}}^{-\alpha}\left|h_{k, {l}^{*}}\right|^{2}\right)\right],
\end{equation}
where $b_{{l}^{*}}$ denotes the closest AP of user $u_{k}$ among ${L}_{m}-{K}_{m}+1$ APs, and a lower-bound of $\bar{R}_{2}$ is given by
\begin{equation}\label{4.1.6}
\bar{R}_{2}\geq\mathbb{E}\left[\log _{2}\left({d}_{k, j^{*}}^{-\alpha}|h_{k, l_{j^{*}}^{*}}|^{2}\right)\right],
\end{equation}
where $u_{{j}^{*}}$ denotes the ${K}_{m}$-th nearest interfering user of user $u_{k}$ and $b_{l_{\hspace{0.5mm}j^{*}}^{*}}$ represents $u_{{j}^{*}}$'s closest AP.
By substituting (\ref{4.1.5}) and (\ref{4.1.6}) into (\ref{4.1.1}),
we have
\begin{equation}\label{4.1.7}
\bar{R}\leq\mathbb{E}\left[\log _{2}\left({d}_{k, {l}^{*}}^{-\alpha}\right)\right]-\mathbb{E}\left[\log _{2}\left({d}_{k, j^{*}}^{-\alpha}\right)\right],
\end{equation}
since the small-scale channel gain $\left|h_{k, {l}^{*}}\right|^{2}$ and $|h_{k, l_{j^{*}}^{*}}|^{2}$ are i.i.d random variables. The average per-user rate $\bar{R}$ can be further obtained as
\begin{align}\label{4.1.8}
\bar{R}&\leq\frac{\alpha}{2}\hspace{-0.5mm}\left[\log _{2}\hspace{-0.5mm}\left(\hspace{-1mm}\left(\lambda\hspace{-0.5mm}-\hspace{-0.5mm}1\right)\hspace{-0.5mm}\frac{K}{M}\hspace{-0.5mm}+\hspace{-0.5mm}1\hspace{-1mm}\right)\hspace{-1mm}-\hspace{-0.5mm}\log _{2}\hspace{-0.5mm}\left(\hspace{-0.5mm}\frac{\lambda K}{L}\right)\hspace{-0.5mm}+\hspace{-0.5mm}\gamma\log _{2}{e}\right],
\end{align}
where $\gamma\approx0.5772$ denotes the Euler-Mascheroni constant. The corresponding derivation is provided in Appendix B.

Since the average energy efficiency $\bar{\eta}_{EE}$ given in (\ref{4.2.3}) increases with the average per-user rate $\bar{R}$ monotonically, an upper-bound of the average energy efficiency, $\bar{\eta}_{EE}^{ub}$, can be obtained by combining (\ref{4.2.3}) and (\ref{4.1.8}) as (\ref{4.2.2}), which is shown at the bottom of this page.
\begin{figure*}[!ht]
	\centering
	\subfloat[$K=100$]{
		\label{1a}
		\includegraphics[width=0.45\textwidth]{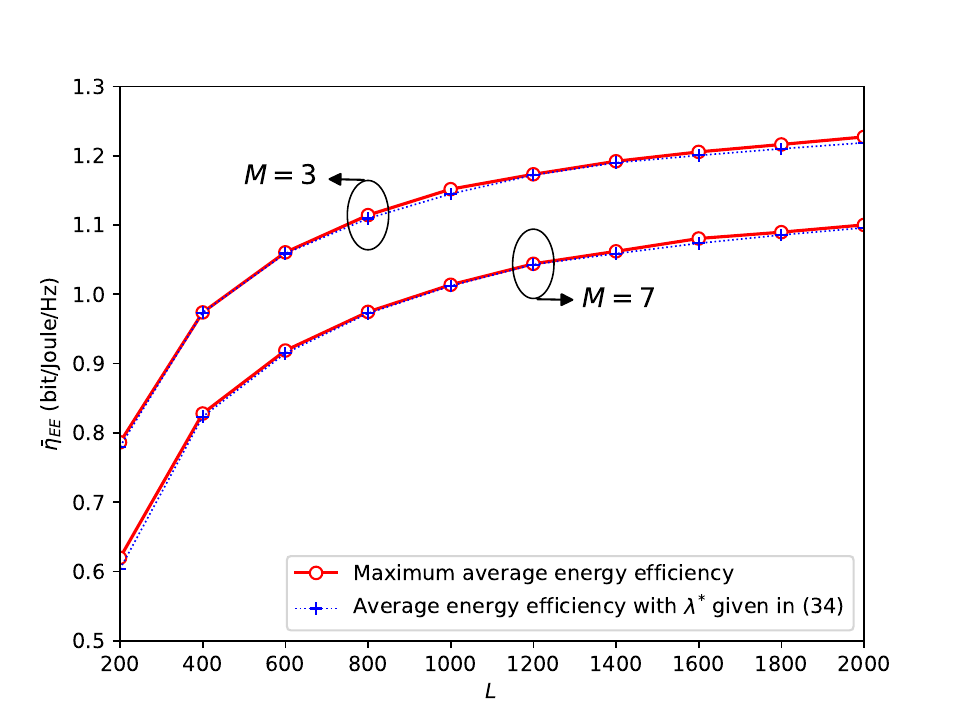}}
    \hspace{8mm}
	\subfloat[$L=1000$]{
		\label{1b}
		\includegraphics[width=0.45\textwidth]{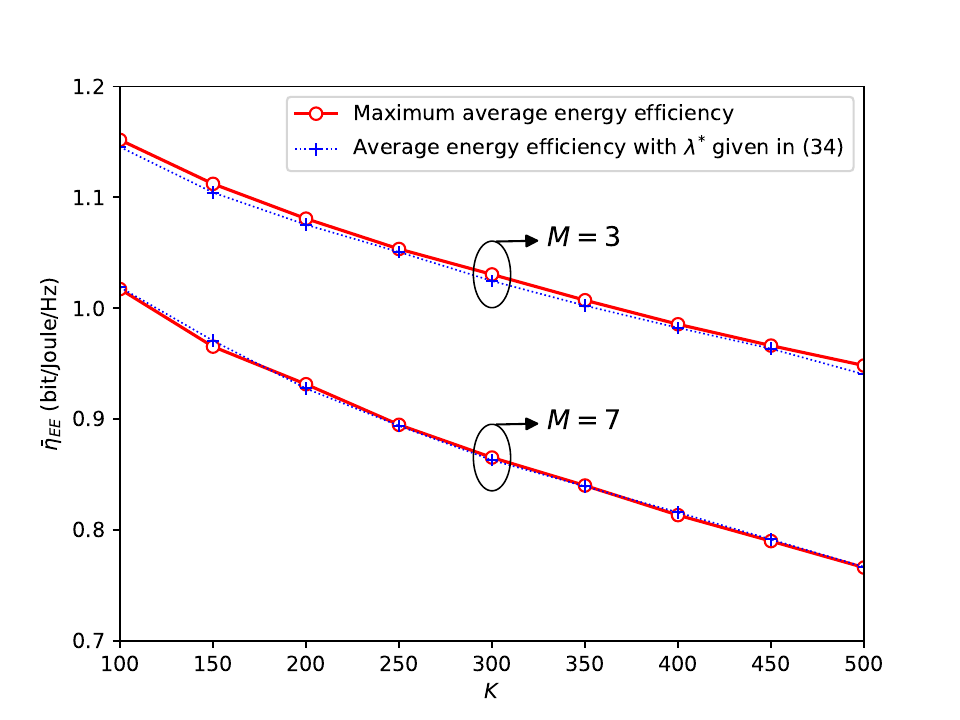}}
	\caption{Maximum average energy efficiency and the average energy efficiency achieved with $\lambda^{*}$ in (\ref{c.3}) versus (a) the total number of APs $L$ and (b) the number of users $K$.}
\end{figure*}
\begin{figure}[t]
	\centering\includegraphics[width=0.45\textwidth]{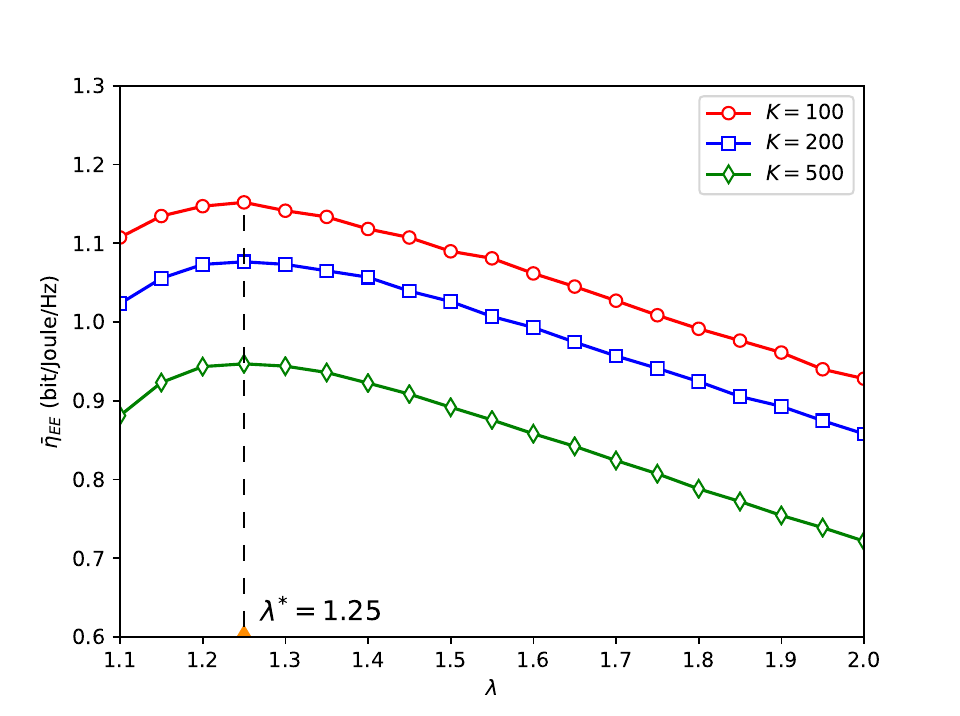}
	\caption{Average energy efficiency $\bar{\eta}_{EE}$ versus the AP-selection ratio $\lambda$. $L=1000$. $M=3$.}
\end{figure}

\subsection{Optimal AP-Selection Ratio $\lambda^{*}$}
Since the per-AP transmission power is usually much larger than the fixed power consumption, 
the upper-bound $\bar{\eta}_{EE}^{ub}$ given in (\ref{4.2.2})
can be approximated as (\ref{4.2.21}), shown at the bottom of this page.
The optimal
AP-selection ratio $\lambda^{*}$ can be then obtained by Theorem 1.

\emph{Theorem 1:
 The optimal AP-selection ratio $\lambda^{*}$ that maximizes the average energy efficiency $\bar{\eta}_{EE}$ follows the solution of 
\begin{equation}
\frac{\partial \bar{\eta}_{EE}^{ub}}{\partial \lambda}\Big|_{\lambda=\lambda^{*}}=0\tag{31}.
\end{equation}
\hspace{1em}Proof: See Appendix C. $\hfill\blacksquare$
}

By taking the first derivative of (\ref{4.2.21}) with respect to $\lambda$,  we have (\ref{c}), shown at the bottom of this page.
According to Theorem 1, one can obtain from (\ref{c}) that
\begin{equation}\label{c.1}
\frac{\lambda^{*}}{\lambda^{*}-1} e^{\frac{\lambda^{*}}{\lambda^{*}-1}}=\frac{L}{M} e^{1+\gamma}\tag{33}.
\end{equation}
The optimal AP-selection ratio $\lambda^{*}$ is then given by \cite{6213038}
\begin{equation}\label{c.3}
\lambda^{*}=\frac{W\left(\frac{L}{M} e^{1+\gamma}\right)}{W\left(\frac{L}{M} e^{1+\gamma}\right)-1}\tag{34},
\end{equation}
where
\begin{equation}\label{c.x}
W(x)\stackrel{x\gg 1}{\approx}\ln x-\ln \ln x+\frac{\ln \ln x}{\ln x}\tag{35}
\end{equation}
denotes the Lambert $W$ function \cite{corless1996lambertw}. Based on (\ref{c.3}), one can easily obtain the optimal number of APs that should be activated, with which the proposed UCR-ApSel algorithm can be quickly run to get the clustered cell-free networking result. This is highly desirable in practical systems in the sense of both reducing the time complexity and quickly adapting to the time-varying environment.

\subsection{Effectiveness of $\lambda^{*}$ Derived in (\ref{c.3})}

Fig. 6 presents the average energy efficiency $\bar{\eta} _{EE}$ versus the AP-selection ratio $\lambda$.
The optimal AP-selection ratio $\lambda^{*}$ given in (\ref{c.3}) is also marked in the figure. It can be observed from
Fig. 6 that (\ref{c.3}) provides a good estimation of the optimal AP-selection ratio that maximizes the average energy efficiency. Additionally, we can see from this figure that 
the optimal AP-selection ratio $\lambda^{*}$ in (\ref{c.3}) is quite accurate when the total number of APs $L$ is large, indicating that the derived optimal AP-selection ratio $\lambda^{*}$ works effectively in future wireless communication systems with ultra-densely deployed APs.

To further verify the effectiveness of the derived AP-selection ratio $\lambda^{*}$ in (\ref{c.3}), Fig. 7 presents the average energy efficiency $\bar{\eta} _{EE}$ achieved with the AP-selection ratio $\lambda^{*}$ given in (\ref{c.3}) and the maximum average energy efficiency achieved by exhaustively searching over the range of $\lambda$ from 1 to $L/K$. As we can clearly observe from Fig. 7 that the average energy efficiency achieved with the optimal $\lambda^{*}$ in (\ref{c.3}) perfectly matches the maximum average energy efficiency, which corroborates that (\ref{c.3}) can accurately estimate the optimal AP-selection ratio $\lambda^{*}$ in the sense of average energy efficiency maximization.

Moreover, it can be observed from (\ref{c.3}) that the optimal AP-selection ratio $\lambda^{*}$ is independent of the number of users $K$.
Fig. 8 plots the average energy efficiency $\bar{\eta}_{EE}$ versus the AP-selection ratio $\lambda$ with 100, 200 and 500 users, respectively.
It can be observed from Fig. 8 that the optimal AP-selection ratio $\lambda^{*}$ remains at 1.25 as 
the number of users $K$ increases from 100 to 500, indicating that the optimal AP-selection ratio $\lambda^{*}$ is 
only determined by the total number of APs $L$ and the number of subnetworks $M$. This indicates that the proposed UCR-ApSel algorithm is robust to the
varying number of users in the network and thus is favourable in practice as the number
of users admitted to the mobile communication system varies over time.

\subsection{Comparison with Benchmarks}
\begin{figure*}[t]
	\centering
	\subfloat[$L=200$]{
		\label{1a}
		\includegraphics[width=0.435\textwidth]{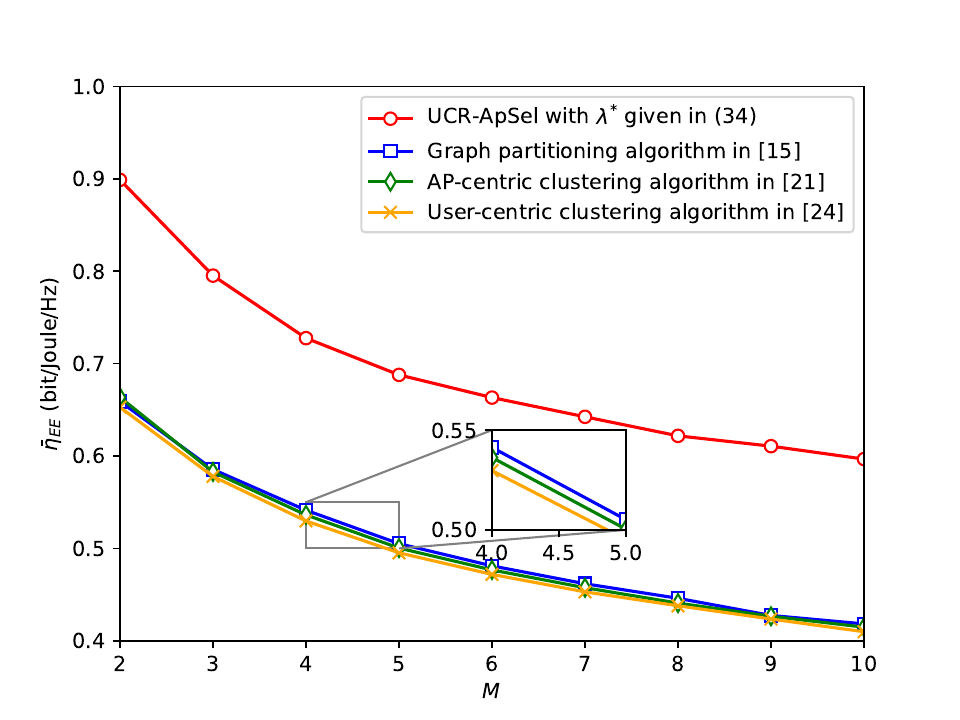}}
      \hspace{8mm}
	\subfloat[$M=3$]{
		\label{1b}
		\includegraphics[width=0.45\textwidth]{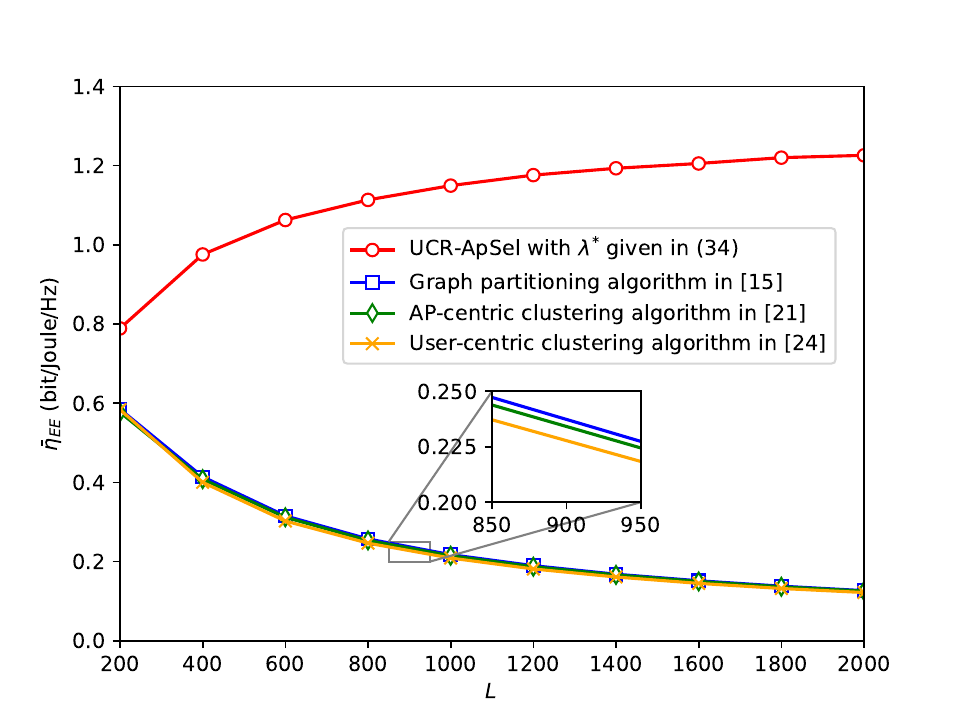}}
	\caption{Average energy efficiency of the proposed UCR-ApSel algorithm with $\lambda^{*}$ in (\ref{c.3}) and the benchmarks versus (a) the number of subnetworks $M$ and (b) the total number of APs $L$. $K=100$.}
\end{figure*}
\begin{figure*}[t]
	\centering
	\subfloat[$M=3$]{
		\label{1a}
		\includegraphics[width=0.435\textwidth]{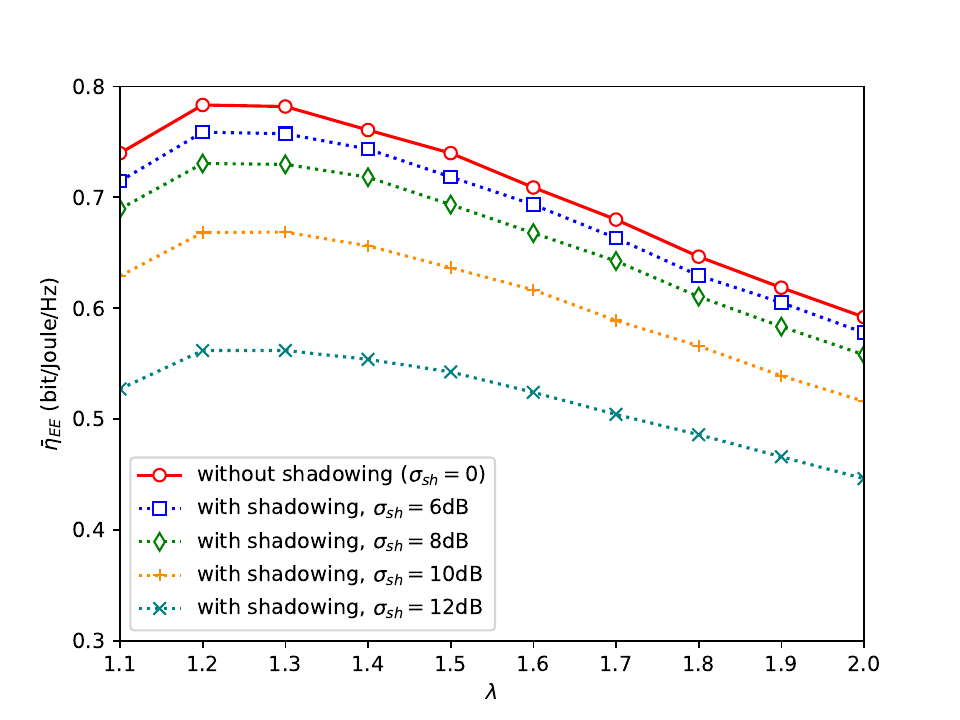}}
  	  \hspace{8mm}
	\subfloat[$M=7$]{
		\label{1b}
   \includegraphics[width=0.45\textwidth]{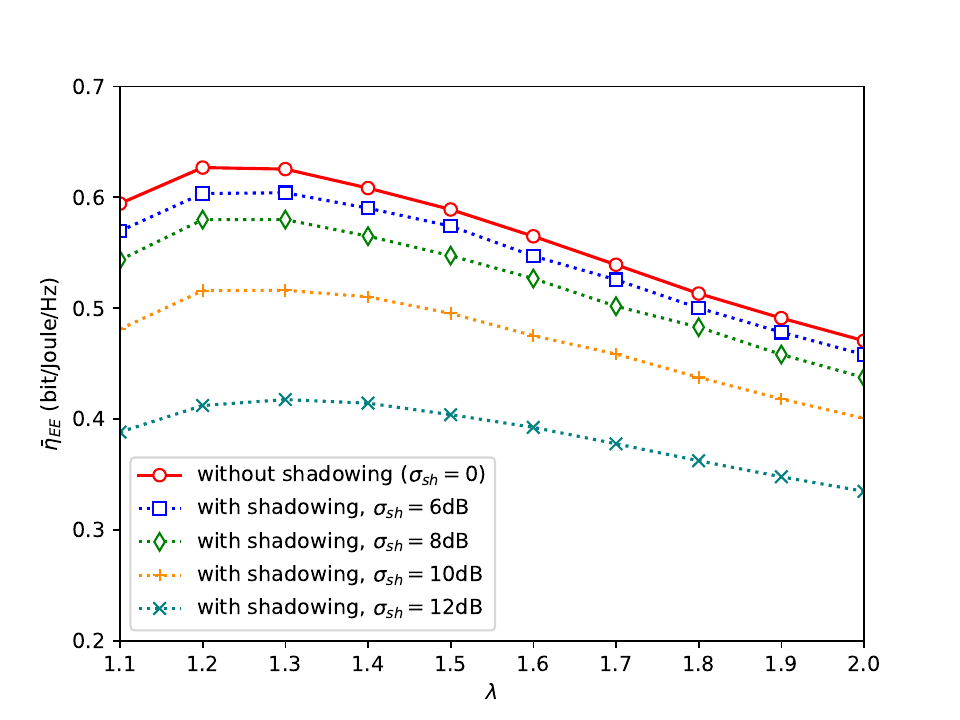}}
		\caption{Average energy efficiency $\bar{\eta} _{EE}$ versus the AP-selection ratio $\lambda$ with the proposed UCR-ApSel algorithm. $L=200$. $K=100$.}
\end{figure*}
To demonstrate the performance gains of the UCR-ApSel algorithm with the optimal AP-selection ratio $\lambda^{*}$, we compare the proposed UCR-ApSel algorithm with 3 existing clustered cell-free networking algorithms:
\begin{itemize}
\item Graph partitioning\cite{JunyuanTWC}: The graph partitioning benchmark in \cite{JunyuanTWC} first merges each user and its closest AP as meganode, and then applies spectral clustering algorithm to partition the whole network into subnetworks.
\item AP-centric\cite{8849663}: The AP-centric benchmark in \cite{8849663} first groups
APs with hierarchical clustering algorithm,  and then each user is affiliated with its nearest AP to form subnetworks.
\item User-centric\cite{8644255}: The User-centric benchmark in \cite{8644255} first clusters users with K-means algorithm, and then each AP is assigned to user clusters with its closest cluster centroid.
\end{itemize}

Fig. 9 presents the average energy efficiency achieved by the proposed UCR-ApSel algorithm with the optimal AP-selection ratio $\lambda^{*}$ and the benchmarks versus the number of subnetworks $M$ and the total number of APs $L$. It can be observed from Fig. 9 that our proposed UCR-ApSel algorithm achieves significant performance gains compared with the baselines thanks to the proper AP selection.  As can be seen in Fig. 9(a), the average energy efficiency $\bar{\eta} _{EE}$ with the proposed UCR-ApSel algorithm is around 40\% higher than that with the baselines when the total number of APs $L=200$.
More importantly, it is shown in Fig. 9(b) that the average energy efficiency of the proposed scheme increases with the total number of APs $L$, while the energy efficiency of the benchmarks decreases. This is due to the fact that the benchmarks activate most or all of the APs for transmission, leading to significant waste of energy, especially with a massive number of APs deployed. The comparison demonstrates the paramount importance of AP selection in future ultra-dense wireless communication systems and the effectiveness of the derived optimal AP-selection ratio $\lambda^{*}$.

\section{Discussion}
Note that the large-scale fading coefficient is modeled by path-loss only in (3) while the shadowing is ignored. To see whether the optimal AP-selection ratio $\lambda^{*}$ that maximizes the average energy efficiency remains the same when shadowing is taken into consideration, let us model the large-scale fading coefficient $\gamma_{k, l}$ from AP $b_{l}$ to user $u_{k}$ with shadowing as
\begin{equation}\label{shadow}
\gamma_{k, l}=\sqrt{\left\|\mathbf{r}_{l}^{B}-\mathbf{r}_{k}^{U}\right\|^{-\alpha}\cdot10^{\frac{c_{k, l}}{10}}}\tag{36},
\end{equation}
where $c_{k, l} \sim \mathcal{N}(0, \sigma_{sh}^{2})$ is a normal random variable with zero mean and standard deviation $\sigma_{sh}$. In the case of no shadowing, i.e., $\sigma_{sh}=0$, (\ref{shadow}) reduces to (3) in the paper, where the large-scale fading coefficient is solely determined by the path-loss. In practice, the standard deviation $\sigma_{sh}$ may vary from 5dB to 12dB \cite{shadow}.

Fig. 10 illustrates the average energy efficiency $\bar{\eta} _{EE}$ versus AP-selection ratio $\lambda$ under different shadowing conditions, i.e., with $\sigma_{sh}$ = 0, 6dB, 8dB, 10dB, 12dB. We can clearly see from this figure that the average energy efficiency varies with the AP-selection ratio in the same manner regardless of $\sigma_{sh}$. More importantly, the optimal AP-selection ratio that maximizes the average energy efficiency seems to be the same under various values of $\sigma_{sh}$. Fig. 10 corroborates that shadowing effect could be ignored when deriving the optimal AP-selection ratio $\lambda^{*}$ for average energy efficiency maximization.
\begin{figure*}[!hb]
	\begin{normalsize}
 		\hrulefill
		\begin{align}\label{a.2}
			\bar{R}_{2}&=\mathbb{E}_{\mathbf{H}, {{\left\{\mathbf{r}_{k}^{U}\right\}_{u_{k} \in \mathcal{U}}}, {\left\{\mathbf{r}_{l}^{B}\right\}_{b_{l} \in \mathcal{B}}}}}\hspace{-2mm}\left\{\hspace{-1mm}\log_{2}\hspace{-1mm}\left(\hspace{-1mm}\sum_{n =1, n \neq m}^{M}\hspace{-0.5mm}\sum_{u_{j} \in \mathcal{U}_{n}}\hspace{-1.5mm}\left[\sum_{b_{l} \in \mathcal{B}_{n}}\hspace{-1mm}\mathrm{g}_{k, l}w_{j, l}{w}_{j, l}^{*}{\mathrm{g}}_{k, l}^{*}\hspace{1mm}+\hspace{-3mm}\sum_{b_{l}, b_{l^{'}} \in \mathcal{B}_{n}, l^{'} \neq l}\hspace{-3mm}\mathrm{g}_{k, l}w_{j, l}{w}_{j, l^{'}}^{*}{\mathrm{g}}_{k, l^{'}}^{*}\hspace{-1mm}\right]\hspace{-1mm}\right)\hspace{-1mm}\right\}\notag \\
			&=\mathbb{E}_{\mathbf{H}, {{\left\{\mathbf{r}_{k}^{U}\right\}_{u_{k} \in \mathcal{U}}}, {\left\{\mathbf{r}_{l}^{B}\right\}_{b_{l} \in \mathcal{B}}}}}\hspace{-2mm}\left\{\log_{2}\left(\sum_{n =1, n \neq m}^{M}\sum_{u_{j} \in \mathcal{U}_{n}}\sum_{b_{l} \in \mathcal{B}_{n}}\left|\mathrm{g}_{k, l}\right|^{2}\cdot\left|w_{j, l}\right|^{2}\right)\right\}
			\notag \\
			&\geq\mathbb{E}_{\mathbf{H}, {{\left\{\mathbf{r}_{k}^{U}\right\}_{u_{k} \in \mathcal{U}}}, {\left\{\mathbf{r}_{l}^{B}\right\}_{b_{l} \in \mathcal{B}}}}}\hspace{-2mm}\left\{\log_{2}\left(\sum_{b_{l} \in \mathcal{B}_{n^{*}}}\left|\mathrm{g}_{k, l}\right|^{2}\cdot\left|w_{j^{*}\hspace{-1mm}, \hspace{0.5mm}l}\right|^{2}\right)\right\}\approx\mathbb{E}\left\{\log_{2}\left({d}_{k, j^{*}}^{-\alpha}|h_{k, l_{j^{*}}^{*}}|^{2}\right)\right\}\tag{A.4}
		\end{align}
   		\hrulefill
		\begin{align}\label{b.1.5}
			\mathbb{E}\left[\log _{2}\left({d}_{k, {l}^{*}}^{-\alpha}\right)\right]&=-\alpha\int_{0}^{{r}_{m}}\log _{2}x\cdot f_{{d}_{k, {l}^{*}}}\left(x\right)dx=-\alpha\left(\left(\lambda-1\right){K}_{m}+1\right)\int_{0}^{{r}_{m}}\left(1-\frac{x^2}{{r}_{m}^2}\right)^{\left(\lambda-1\right){K}_{m}}\left(\frac{2x}{{r}_{m}^2}\right)\log _{2}xdx\notag \\
			&\stackrel{y=\frac{x^2}{{r}_{m}^{2}}}=-\frac{\alpha}{2}\left(\left(\lambda-1\right){K}_{m}+1\right)\int_{0}^{1}\left(1-y\right)^{\left(\lambda-1\right){K}_{m}}\log _{2}ydy-\frac{\alpha}{2}\log _{2}\left(\frac{\lambda{K}_{m}D^{2}}{L}\right)\notag \\
			&=\frac{\alpha}{2}\left(H_{\left(\lambda-1\right){K}_{m}+1}\log _{2}{e}-\log _{2}\left(\frac{\lambda{K}_{m}D^{2}}{L}\right)\right)\tag{B.5}
		\end{align}
	\end{normalsize}
\end{figure*}

\section{Conclusion}
This paper investigated the energy-efficient clustered cell-free networking problem with AP selection for future ultra-dense wireless networks with massive distributed APs and users. We proposed a novel UCR-ApSel algorithm to incorporate AP selection into clustered cell-free networking for energy saving.
Simulation results showed that our proposed UCR-ApSel algorithm achieves higher average energy efficiency and better user fairness than the benchmark.
We then analyzed the average energy efficiency with the proposed UCR-ApSel algorithm and a closed-form upper-bound was obtained. Based on the theoretical upper-bound, the optimal AP-selection ratio was derived as an explicit function of the total number of APs $L$ and the number of subnetworks $M$, thanks to which, the optimal number of APs that should be activated can be quickly determined in practice to facilitate adaptive clustered cell-free networking in dynamic environments.

Note that equal power allocation over users is assumed
in this paper as a starting point, while the potential of clustered cell-free networks can be
further exploited by jointly optimizing power allocation and clustered cell-free networking
along with AP selection, which will be carefully studied in our future work. Moreover,
the proposed UCR-ApSel algorithm is performed in a centralized manner, which might be unaffordable in future ultra-dense wireless  networks. How to design
a distributed clustered cell-free networking scheme is an interesting topic, which deserves
much attention in future study.

\numberwithin{equation}{section}
\begin{appendices}
\section{Derivation of (\ref{4.1.5}) and (\ref{4.1.6})}
According to (\ref{4.1.1}), the average per-user rate $\bar{R}$ is determined by
\begin{equation}\bar{R}_{1}\hspace{-1mm}=\hspace{-0.5mm}\mathbb{E}_{\mathbf{H}, {{\left\{\hspace{-0.5mm}\mathbf{r}_{k}^{U}\hspace{-0.5mm}\right\}_{u_{k} \in \mathcal{U}}}, {\left\{\hspace{-0.5mm}\mathbf{r}_{l}^{B}\hspace{-0.5mm}\right\}_{b_{l} \in \mathcal{B}}}}}\hspace{-1mm}\left[\log_{2}\hspace{-1mm}\left(\hspace{-0.5mm}\mathbf{g}_{k, \mathcal{B}_{m}}\hspace{-0.5mm} \mathbf{w}_{k, \mathcal{B}_{m}}\hspace{-0.5mm}\mathbf{w}_{k, \mathcal{B}_{m}}^{\dagger}\hspace{-0.5mm} \mathbf{g}_{k, \mathcal{B}_{m}}^{\dagger}\hspace{-0.5mm}\right)\hspace{-1mm}\right],
\end{equation}
and
\begin{equation}\label{4.1.3}\bar{R}_{2}\hspace{-1.25mm}=\hspace{-1mm}\mathbb{E}_{\hspace{-0.125mm}\mathbf{H}\hspace{-0.25mm}, \hspace{-0.5mm}{{\left\{\hspace{-0.5mm}\mathbf{r}_{k}^{U}\hspace{-0.75mm}\right\}_{\hspace{-0.75mm}u_{\hspace{-0.25mm}k}\hspace{-0.5mm} \in\hspace{-0.125mm} \mathcal{U}}}\hspace{-0.5mm},\hspace{-0.5mm} {\left\{\hspace{-0.5mm}\mathbf{r}_{l}^{B}\hspace{-0.75mm}\right\}_{\hspace{-0.75mm}b_{\hspace{-0.125mm}l}\hspace{-0.5mm} \in\hspace{-0.25mm} \mathcal{B}}}}}\hspace{-1.75mm}\left[\hspace{-0.75mm}\log _{\hspace{-0.25mm}2}\hspace{-1.5mm}\left(\hspace{-0.25mm}\sum_{\substack{n =1\\n \neq m}}^{M}\hspace{-0.75mm}\sum_{u_{j} \in \mathcal{U}_{n}}\hspace{-2mm}\mathbf{g}_{k\hspace{-0.25mm}, \mathcal{B}_{n}}\hspace{-0.75mm}\mathbf{w}_{j\hspace{-0.25mm}, \mathcal{B}_{n}}\hspace{-0.75mm}\mathbf{w}_{j\hspace{-0.25mm}, \mathcal{B}_{n}}^{\dagger}\hspace{-0.75mm}\mathbf{g}_{k\hspace{-0.25mm}, \mathcal{B}_{n}}^{\dagger}\hspace{-1.5mm}\right)\hspace{-1.25mm}\right].
\end{equation}
With ZFBF performed in each subnetwork,
$\bar{R}_{1}$ can be approximated as \cite{7031453}
\begin{equation}\label{a.1}
\bar{R}_{1}\hspace{-1mm}\approx\hspace{-0.75mm}\mathbb{E}\hspace{-1.25mm}\left[\hspace{-0.5mm}\log _{2}\hspace{-1mm}\left(\hspace{-0.25mm}\sum_{l \in \tilde{\mathcal{B}}_{m}}\hspace{-1mm}d_{k, l}^{-\alpha}\hspace{-0.5mm}\left|h_{k, l}\right|^{2}\hspace{-1mm}\right)\hspace{-1mm}\right]\hspace{-1mm}\stackrel{\text{for large}\hspace{1mm}L}{\approx}\hspace{-0.5mm}\mathbb{E}\hspace{-0.5mm}\left[\hspace{-0.25mm}\log _{2}\hspace{-0.75mm}\left(\hspace{-0.5mm}{d}_{k, {l}^{*}}^{-\alpha}\hspace{-1mm}\left|h_{k, {l}^{*}}\hspace{-0.5mm}\right|^{2}\hspace{-0.5mm}\right)\hspace{-0.5mm}\right],
\end{equation}
where $\tilde{\mathcal{B}}_{m}$ denotes the set of ${L}_{m}-{K}_{m}+1$ APs and 
${d}_{k, {l}^{*}}$ represents the minimum distance between user $u_{k}$ and the APs in $\tilde{\mathcal{B}}_{m}$.
$\bar{R}_{2}$ can be obtained by combining (\ref{2.1.2}), (\ref{2.1.3}) and (\ref{4.1.3}) as (\ref{a.2}), which is shown at the bottom of this page,
where $u_{{j}^{*}}$ in the $n^{*}$th subnetwork denotes the ${K}_{m}$-th nearest interfering user of user $u_{k}$ and $b_{l_{j^{*}}^{*}}$ denotes $u_{{j}^{*}}$'s closest AP.
(\ref{4.1.6}) can be then easily derived from (\ref{a.2}).
\begin{figure*}[!ht]
	\begin{normalsize}
		\begin{align}\label{b.2.2}
			\mathbb{E}\hspace{-1mm}\left[\log _{2}\hspace{-0.75mm}\left(\hspace{-0.75mm}{d}_{k, j^{*}}^{-\alpha}\hspace{-1mm}\right)\hspace{-0.5mm}\right]&\hspace{-1.5mm}=\hspace{-0.75mm}-\alpha\hspace{-1mm}\int_{0}^{D}\hspace{-1mm}\log _{2}x\hspace{-0.5mm}\cdot\hspace{-0.5mm} f_{{d}_{k, {j}^{*}}}\hspace{-1mm}\left(x\right)dx\hspace{-0.5mm}=\hspace{-0.5mm}\frac{-\alpha\left(K-1\right)!}{\left({K}_{m}-1\right)!\left(K-1-{K}_{m}\right)!}\int_{0}^{D}\hspace{-1.5mm}\left(\frac{x}{D}\right)^{\hspace{-0.5mm}2\left({K}_{m}-1\right)}\hspace{-1mm}\left(\hspace{-0.5mm}1\hspace{-0.5mm}-\hspace{-0.5mm}\frac{x^{2}}{D^{2}}\right)^{K-1-{K}_{m}}\hspace{-0.5mm}\left(\frac{2x}{D^{2}}\right)\log _{2}xdx\notag \\
			&\hspace{-4mm}\stackrel{y=\frac{x^{2}}{D^{2}}}=\hspace{-2mm}\frac{\left(K-1\right)!}{\left(\hspace{-0.5mm}{K}_{m}\hspace{-1mm}-\hspace{-1mm}1\hspace{-0.5mm}\right)!\hspace{-0.5mm}\left(\hspace{-0.5mm}K\hspace{-1mm}-\hspace{-1mm}1\hspace{-1mm}-\hspace{-1mm}{K}_{m}\hspace{-0.5mm}\right)!}\hspace{-0.75mm}\left(\hspace{-1.25mm}-\frac{\alpha}{2}\hspace{-1mm}\int_{0}^{1}\hspace{-1.5mm}y^{{K}_{\hspace{-0.25mm}m}\hspace{-0.5mm}-\hspace{-0.25mm}1}\hspace{-0.75mm}\left(\hspace{-0.25mm}1\hspace{-0.5mm}-\hspace{-0.5mm}y\hspace{-0.25mm}\right)^{\hspace{-0.25mm}K\hspace{-0.25mm}-\hspace{-0.25mm}1-\hspace{-0.25mm}{K}_{\hspace{-0.25mm}m}}\hspace{-0.5mm}\log _{2}\hspace{-0.25mm}ydy\hspace{-0.75mm}\right)\hspace{-1.25mm}-\hspace{-0.75mm}\alpha\hspace{-0.25mm}\log _{2}\hspace{-0.25mm}D=-\frac{\alpha\left(H_{{K}_{m}-1}-H_{K-1}\right)}{2\ln2}-\alpha\log _{2}D\tag{B.9}
		\end{align}
     		\hrulefill
	\end{normalsize}
\end{figure*}

\section{Derivation of (\ref{4.1.8})}
Recall that it is given in (\ref{4.1.7}) that the average per-user rate $\bar{R}\leq\mathbb{E}\left[\log _{2}\left({d}_{k, {l}^{*}}^{-\alpha}\right)\right]-\mathbb{E}\left[\log _{2}\left({d}_{k, j^{*}}^{-\alpha}\right)\right]$. To analyze the average per user rate $\bar{R}$,
let us first take a closer look at
$\mathbb{E}\left[\log _{2}\left({d}_{k, {l}^{*}}^{-\alpha}\right)\right]$ and 
$\mathbb{E}\left[\log _{2}\left({d}_{k, {j}^{*}}^{-\alpha}\right)\right]$, respectively.

\subsection{$\mathbb{E}\left[\log _{2}\left({d}_{k, {l}^{*}}^{-\alpha}\right)\right]$}
With the proposed UCR-ApSel algorithm, in the $m$th subnetwork,
each of the ${L}_{m}$ APs in $\mathcal{B}_{m}$ is located close to some user in $\mathcal{U}_{m}$.
The distance ${d}_{k, {l}^{*}}$ between reference user $u_{k}$ and its nearest AP $b_{{l}^{*}}$ among ${L}_{m}-{K}_{m}+1$ APs
could be approximated by the minimum distance between user $u_{k}$ and ${L}_{m}-{K}_{m}+1$ randomly distributed APs within a circular area
of radius ${r}_{m}$, given by
\begin{equation}\label{b.1.1}
{r}_{m}\approx\sqrt{\frac{{L}_{m}}{L}}D.
\end{equation}
Without loss of generality, by assuming that user $u_{k}$ is located at the center of the circular area, the minimum distance ${d}_{k, {l}^{*}}$ follows the pdf
\begin{equation}\label{b.1.2}
f_{{d}_{k, {l}^{*}}}\hspace{-1mm}\left(x)\hspace{-0.5mm}=\hspace{-0.5mm}({L}_{m}-{K}_{m}+1\right)\hspace{-0.5mm}\left(1-F\left(x;{r}_{m}\right)\right)^{{L}_{m}-{K}_{m}}\hspace{-0.5mm}f\hspace{-0.5mm}\left(x;{r}_{m}\hspace{-0.25mm}\right),
\end{equation}
where
\begin{equation}\label{b.1.3}
f\left(x; y\right)=\frac{2x}{y^2} \quad 0 \leq x \leq {y},
\end{equation}
and
\begin{equation}\label{b.1.4}
F\left(x; y\right)=\frac{x^2}{y^2} \quad 0 \leq x \leq {y}.
\end{equation}
$\mathbb{E}\left[\log _{2}\left({d}_{k, {l}^{*}}^{-\alpha}\right)\right]$ can be then obtained by combining (\ref{b.1.1})-(\ref{b.1.4}) as (\ref{b.1.5}), which is shown at the bottom of the previous page, where $ H_{n}=\sum_{i=1}^{n}1/i$ denotes the Harmonic series.
When the number of users in each subnetwork is much larger than 1, i.e., ${K}_{m}\gg 1$, we have
\begin{equation}\label{b.1.88}
H_{\left(\lambda-1\right){K}_{m}+1}\approx \ln\left(\left(\lambda-1\right){K}_{m}+1\right)+\gamma\tag{B.6},
\end{equation}
where $\gamma\approx0.5772$ denotes the Euler-Mascheroni constant. Since
balanced user groups would be generated by partitioning uniformly
distributed users with agglomerative hierarchical clustering, i.e., ${K}_{m}\approx K/M$, $\forall m$, it can be obtained from (\ref{b.1.5}) and (\ref{b.1.88}) that 
\begin{equation}\label{b.1.6}\mathbb{E}\hspace{-1mm}\left[\hspace{-0.5mm}\log _{2}\hspace{-1mm}\left(\hspace{-1mm}{d}_{k, {l}^{*}}^{-\alpha}\hspace{-1mm}\right)\hspace{-0.75mm}\right]\hspace{-1.25mm}\approx\hspace{-1mm}\frac{\alpha}{2}\hspace{-1mm}\left[\hspace{-0.5mm}\log _{2}\hspace{-1mm}\left(\hspace{-1.5mm}\left(\hspace{-0.5mm}\lambda\hspace{-0.75mm}-\hspace{-0.75mm}1\hspace{-0.25mm}\right)\hspace{-0.75mm}\frac{K}{M}\hspace{-0.75mm}+\hspace{-1mm}1\hspace{-1.25mm}\right)\hspace{-1.25mm}-\hspace{-0.75mm}\log _{2}\hspace{-1mm}\left(\hspace{-1mm}\frac{\lambda KD^{2}}{LM}\hspace{-0.75mm}\right)\hspace{-1.25mm}+\hspace{-0.75mm}\gamma\hspace{-0.5mm}\log _{2}{\hspace{-0.5mm}e}\hspace{-0.5mm}\right]\tag{B.7}.
\end{equation}

\subsection{$\mathbb{E}\left[\log _{2}\left({d}_{k, {j}^{*}}^{-\alpha}\right)\right]$}
As the users are uniformly distributed within a circular area with radius $D$, the distance between reference user $u_{k}$ and its ${K}_{m}$-th
closest user $u_{{j}^{*}}$, ${d}_{k, j^{*}}$, follows the pdf \cite{david2004order}
\begin{align}\label{b.2.1}
f_{{d}_{k, j^{*}}}\left(x\right)=&\frac{\left(K-1\right)!}{\left({K}_{m}-1\right)!\left(K-1-{K}_{m}\right)!}F\left(x; D\right)^{{K}_{m}-1}\notag \\
&\cdot\left(1-F\left(x; D\right)\right)^{K-1-{K}_{m}}f\left(x; D\right)\tag{B.8}.
\end{align}
$\mathbb{E}\left[\log _{2}\left({d}_{k, {j}^{*}}^{-\alpha}\right)\right]$ can be then obtained from (\ref{b.2.1}) as (\ref{b.2.2}), shown at the top of this page. It can be seen from (\ref{b.1.88}) that $H_{{K}_{m}-1}\approx \ln\left({K}_{m}-1\right)+\gamma$ and 
$H_{K-1}\approx \ln\left(K-1\right)+\gamma$. One can then obtain $H_{{K}_{m}-1}-H_{K-1}\approx-\ln M$. As a result, (\ref{b.2.2}) can be reduced to
\begin{equation}\label{b.2.3}
\mathbb{E}\left[\log _{2}\left({d}_{k, j^{*}}^{-\alpha}\right)\right]=\frac{\alpha}{2}\log _{2} M-\alpha\log _{2}D\tag{B.10}.
\end{equation}
Finally, (\ref{4.1.8}) can be obtained by substituting (\ref{b.1.6}) and (\ref{b.2.3}) into (\ref{4.1.7}).

\section{Proof of Theorem 1}
\emph{Proof}: The closed-form upper-bound of average energy efficiency, $\bar{\eta}_{EE}^{ub}$, given in (\ref{4.2.21}) is a function of the AP-selection ratio $\lambda$, which can be rewritten as
\begin{equation}\label{c.15}
\bar{\eta}_{EE}^{ub}=\frac{f\left(\lambda\right)}{P_{t}\lambda+P_{b} f\left(\lambda\right)},
\end{equation}
where $P_{t}=\frac{P}{\tau}+P_{c}+P_{fix}$ and
\begin{equation}
f\left(\lambda\right)=\frac{\alpha}{2}\left(\log _{2}\left(\frac{\lambda-1}{\lambda}\right)+\log _{2}\left(\frac{L}{M}\right)+\gamma\log _{2}{e}\right).
\end{equation}
The first derivative of $\bar{\eta}_{EE}^{ub}$ with respect to $\lambda$ can be obtained from (\ref{c.15}) as
\begin{equation}\label{c.11}
\frac{\partial \bar{\eta}_{EE}^{ub}}{\partial \lambda}=\frac{P_{t}\mathit{\Phi}\left(\lambda\right)}{\left(P_{t}\lambda+P_{b} f\left(\lambda\right)\right)^{2}},
\end{equation}
where
\begin{equation}\label{c.4}\mathit{\Phi}\left(\lambda\right)\hspace{-0.5mm}=\hspace{-0.5mm}\frac{\alpha}{2}\hspace{-0.5mm}\left(\frac{\log _{2}{e}}{\lambda-1}\hspace{-0.5mm}+\hspace{-0.5mm}\log _{2}\hspace{-0.5mm}\left(\hspace{-0.5mm}\frac{\lambda}{\lambda-1}\hspace{-0.5mm}\right)\hspace{-1mm}-\hspace{-0.5mm}\log _{2}\hspace{-0.5mm}\left(\frac{L}{M}\right)\hspace{-1mm}-\hspace{-0.5mm}\gamma\log _{2}{\hspace{-0.5mm}e}\hspace{-1mm}\right).
\end{equation}
Note that the AP-selection ratio $\lambda$ varies between $1$ and $L/K$, i.e., 
$1<\lambda\leq L/K$, we have
${\lim_{\lambda\to 1^{+}}\mathit{\Phi}\left(\lambda\right)>0}$ and ${\lim_{L/K\to +\infty}\mathit{\Phi}\left(L/K\right)<0}$.
Moreover, the derivative of  $\mathit{\Phi}\left(\lambda\right)$ can be derived from (\ref{c.4}) as
\begin{equation}\label{c.12}
\frac{\partial \mathit{\Phi}\left(\lambda\right)}{\partial \lambda}=\frac{\alpha\log _{2}{e}}{2}\cdot\frac{1-2\lambda}{\lambda\left(\lambda-1\right)^{2}}<0.
\end{equation}
It can be seen from the above that as the AP-selection ratio $\lambda$ increases from $1$ to $L/K$, $\mathit{\Phi}\left(\lambda\right)$ decreases monotonically from some positive value to a negative one. As a result, the derivative $\frac{\partial \bar{\eta}_{EE}^{ub}}{\partial \lambda}>0$ when the AP-selection ratio $\lambda$ is smaller than some constant $\lambda^{*}$, and the derivative $\frac{\partial \bar{\eta}_{EE}^{ub}}{\partial \lambda}<0$ when $\lambda>\lambda^{*}$. Therefore, the maximum average energy efficiency is achieved when $\lambda=\lambda^{*}$, i.e., the optimal AP-selection ratio $\lambda^{*}$ follows the solution of $\frac{\partial \bar{\eta}_{EE}^{ub}}{\partial \lambda}\Big|_{\lambda=\lambda^{*}}=0$.

\end{appendices}

\bibliographystyle{IEEEtran}
\bibliography{reference}

\begin{IEEEbiography}[{\includegraphics[width=1in,height=1.25in,clip,keepaspectratio]{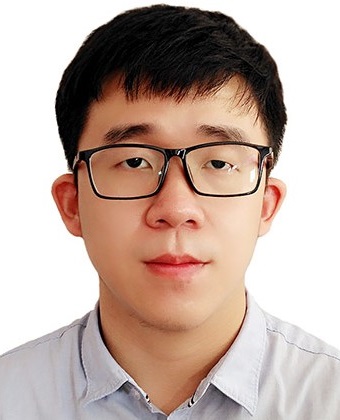}}]{Ouyang Zhou}\enspace received the B.S. degree from Nankai University, Tianjin, China, in 2019. He is currently pursuing the Ph.D. degree with the College of Electronic and Information Engineering, Tongji University, Shanghai China. His current research interests include clustered cell-free networking and deep reinforcement learning. 
\end{IEEEbiography}

\begin{IEEEbiography}[{\includegraphics[width=1in,height=1.25in,clip,keepaspectratio]{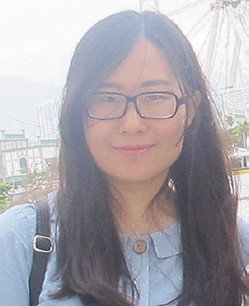}}]{Junyuan Wang}\enspace (Member, IEEE) received the B.S. degree in Communications Engineering from Xidian University, Xi'an, China, in 2010, and the Ph.D. degree in Electronic Engineering from City University of Hong Kong, Hong Kong, China, in 2015. From 2015 to 2017, she was a Research Associate in the School of Engineering and Digital Arts, University of Kent, Canterbury, U.K. From 2018 to 2020, she was a Lecturer (Assistant Professor) in the Department of Computer Science, Edge Hill University, Ormskirk, U.K. She is currently a Research Professor with the College of Electronic and Information Engineering and the Institute for Advanced Study, Tongji University, Shanghai, China. Her research mainly focuses on wireless communications and networking, and  artificial intelligence. Dr. Wang was a co-recipient of the Best Student Paper Award at the IEEE 85th Vehicular Technology Conference–Spring 2017, and a recipient of the Shanghai Leading Talent Program (Young Scientist) in 2021.
\end{IEEEbiography}

\begin{IEEEbiography}[{\includegraphics[width=1in,height=1.25in,clip,keepaspectratio]{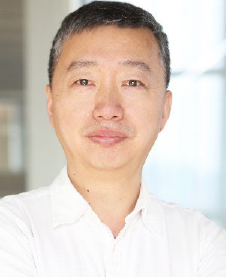}}]{Fuqiang Liu}\enspace (Member, IEEE) received the bachelor's degree from Tianjin University, Tianjin, China, in 1987, and the Ph.D. degree from the China University of Mining and Technology, Xuzhou, China, in 1996. In 2005, he was a Visiting Scholar with University Erlangen-Nürnberg, Erlangen, Germany. He is currently a Professor with the College of Electronics and Information Engineering, Tongji University, Shanghai, China, and also with the College of Computer Science and Technology, Huaibei Normal University, Huaibei, China. He also serves as the Director of the Broadband Wireless Communication and Artificial Intelligence Laboratory, Tongji University. His research interests include information and communications technologies and innovation applications in automotive and intelligent transportation systems.
\end{IEEEbiography}

\newpage 

\begin{IEEEbiography}[{\includegraphics[width=1in,height=1.25in,clip,keepaspectratio]{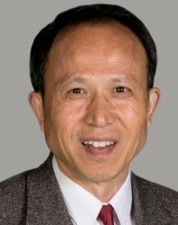}}]{Jiangzhou Wang}\enspace (Fellow, IEEE) is a Professor with the University of Kent, U.K. He has published more than 500 papers and five books. His research focuses on mobile communications. He was a recipient of the 2022 IEEE Communications Society Leonard G. Abraham Prize. He was the Technical Program Chair of the 2019 IEEE International Conference on Communications, Shanghai, the Executive Chair of the IEEE ICC2015, London, and the Technical Program Chair of the IEEE WCNC2013. He has served as an Editor for a number of international journals, including IEEE Transactions on Communications from 1998 to 2013. Professor Wang is a Foreign Member of the Chinese Academy of Engineering (CAE), a Fellow of the Royal Academy of Engineering (RAEng), U.K., Fellow of the IEEE, and Fellow of the IET.
\end{IEEEbiography}

\end{document}